\documentclass[review]{elsarticle}
\usepackage[margin=3cm]{geometry}

\usepackage{lineno,hyperref}

\usepackage{caption}
\usepackage{subcaption}
\usepackage{acro}
\acsetup{first-style=short}
\usepackage{dblfloatfix}
\usepackage{amsmath}
\usepackage[table]{xcolor}
\usepackage{graphicx}
\graphicspath{{./figures/}}
\usepackage[outdir=./]{epstopdf}
\usepackage{booktabs}
\usepackage{siunitx}
\usepackage[utf8]{inputenc}
\usepackage{upquote}
\usepackage{csquotes}

\usepackage{newtxmath}

\modulolinenumbers[5]

\journal{International Journal of Multiphase Flow}

\biboptions{authoryear}

\newcommand{\reff}[1]{Fig.~\ref{#1}}
\newcommand{\refe}[1]{Eq.~\eqref{#1}}
\newcommand{\refs}[1]{Sec.~\ref{#1}}
\newcommand{\reft}[1]{Table~\ref{#1}}

\newcommand{\OMEa}{\mathrm{OME_1}}

\newcommand{\Pl}{P_{\mathrm{l}}}

\newcommand{\Tvplus}{T_{\mathrm{v}}^{+}}
\newcommand{\Pgplus}{P_{\mathrm{g}}^{+}}
\newcommand{\Pv}{P_{\mathrm{v}}}
\newcommand{\Pvplus}{P_{\mathrm{v}}^{+}}
\newcommand{\Pplus}{P^{+}}
\newcommand{\Pplusprime}{P^{+\prime}}
\newcommand{\rhol}{\rho_{\mathrm{l}}}
\newcommand{\rholzero}{\rho_{\mathrm{l0}}}
\newcommand{\rholplus}{\rho_{\mathrm{l}}^{+}}
\newcommand{\rhov}{\rho_{\mathrm{v}}}
\newcommand{\alphal}{\alpha_{\mathrm{l}}}

\newcommand{\cl}{c_{\mathrm{l}}}

\newcommand{\Tsat}{T_{\mathrm{sat}}}
\newcommand{\Tl}{T_{\mathrm{l}}}
\newcommand{\Td}{T_{\mathrm{d}}}
\newcommand{\Tdplus}{T^+_{\mathrm{d}}}
\newcommand{\Tb}{T_{\mathrm{b}}}
\newcommand{\Tv}{T_{\mathrm{v}}}
\newcommand{\Lv}{L_{\mathrm{v}}}
\newcommand{\Rc}{R_{\mathrm{c}}}
\newcommand{\Rref}{R_{\mathrm{ref}}}
\newcommand{\tref}{t_{\mathrm{ref}}}
\newcommand{\Rbstar}{R_\mathrm{b}^{*}}
\newcommand{\tstar}{t^{*}}
\newcommand{\lambdal}{\lambda_{\mathrm{l}}}

\newcommand{\Phd}{P_{\mathrm{hd}}}
\newcommand{\Phdprime}{P_{\mathrm{hd}}^\prime}
\newcommand{\Phdplus}{P_{\mathrm{hd}}^{+}}
\newcommand{\Phdplusprime}{P_{\mathrm{hd}}^{+\prime}}
\newcommand{\Phdaccbplus}{P_{\mathrm{hd,acc}}^{+}}
\newcommand{\Phdaccbplusprime}{P_{\mathrm{hd,acc}}^{+\prime}}

\newcommand{\Phdvelplusprime}{P_{\mathrm{hd,vel}}^{+\prime}}
\newcommand{\Phdaccb}{P_{\mathrm{hd,acc}}}
\newcommand{\Phdaccbprime}{P_{\mathrm{hd,acc}}^\prime}

\newcommand{\Phdvelb}{P_{\mathrm{hd,vel,b}}}
\newcommand{\Phdvelbprime}{P_{\mathrm{hd,vel,b}}^\prime}

\newcommand{\Phdveldbprime}{P_{\mathrm{hd,vel,db}}^\prime}
\newcommand{\Phdvelbplus}{P_{\mathrm{hd,vel,b}}^{+}}
\newcommand{\Phdvelbplusprime}{P_{\mathrm{hd,vel,b}}^{+\prime}}

\newcommand{\Phdveldbplusprime}{P_{\mathrm{hd,vel,db}}^{+\prime}}

\newcommand{\Rb}{R_\mathrm{b}}
\newcommand{\Rd}{R_\mathrm{d}}
\newcommand{\Rbplus}{R_\mathrm{b}^{+}}
\newcommand{\Rdplus}{R_\mathrm{d}^{+}}
\newcommand{\Rdplustwo}{R_\mathrm{d}^{+2}}
\newcommand{\Rbplustwo}{R_\mathrm{b}^{+2}}
\newcommand{\Ndenplus}{n^{+}}
\newcommand{\Ri}{r_{i}}
\newcommand{\Pg}{P_\mathrm{g}}
\newcommand{\Pinertia}{P_\mathrm{inter}}

\newcommand{\ul}{u_\mathrm{l}}

\newcommand{\mul}{\mu_{\mathrm{l}}}
\newcommand{\mulzero}{\mu_{\mathrm{l0}}}

\newcommand{\mulplus}{\mu_{\mathrm{l}}^{+}}
\newcommand{\sigmaplus}{\sigma^{+}}
\newcommand{\sigmazero}{\sigma_{0}}

\newcommand{\Pmu}{P_{\upmu}}
\newcommand{\Pmuprime}{P_{\upmu}^\prime}
\newcommand{\Pmuplus}{P_{\upmu}^{+}}
\newcommand{\Pmuplusprime}{P_{\upmu}^{+\prime}}
\newcommand{\Psigma}{P_{\upsigma}}
\newcommand{\Psigmaprime}{P_{\upsigma}^\prime}
\newcommand{\Psigmaplus}{P_{\upsigma}^{+}}
\newcommand{\Psigmaplusprime}{P_{\upsigma}^{+\prime}}

\newcommand{\DelP}{\Delta P}
\newcommand{\DelPprime}{\Delta P^\prime}
\newcommand{\DelPplus}{\Delta P^{+}}
\newcommand{\DelPplusprime}{\Delta P^{+\prime}}
\newcommand{\DeltaTplus}{\Delta T^+}
\newcommand{\Rgas}{\mathcal{R}}
\newcommand{\Rplus}{R^{+}}
\newcommand{\tplus}{t^{+}}
\newcommand{\tplustwo}{t^{+2}}

\newcommand{\Tnm}{t_\text{num}}
\newcommand{\Tan}{t_\text{anlt}}

\begin{document}
\begin{frontmatter}

\title{{Dimensional Analysis of Vapor Bubble Growth Considering Bubble-bubble Interactions in Flash Boiling Microdroplets of Highly Volatile Liquid Electrofuels}}

%
%
%

\author[mymainaddress]{A. Saha\corref{mycorrespondingauthor}}
\cortext[mycorrespondingauthor]{Corresponding author}
\ead{a.saha@itv.rwth-aachen.de}
\author[mymainaddress]{A. Y. Deshmukh}
\author[mymainaddress,mysecondaryaddress]{T. Grenga}
\author[mymainaddress]{H. Pitsch}

\address[mymainaddress]{Institute for Combustion Technology, RWTH Aachen University, Aachen, Germany}
\address[mysecondaryaddress]{Faculty of Engineering and Physical Science, University of Southampton, Southampton, United Kingdom}

\begin{abstract}
Electrofuels (e-fuels) produced from renewable electricity and carbon sources have gained significant attention in recent years as promising alternatives to fossil fuels for the transportation sector. However, the highly volatile e-fuels, such as short-chain oxymethylene ethers ($\mathrm{OME_x}$) are prone to flash vaporization phenomena, which is associated with the formation and growth of vapor bubbles, followed by explosive bursting of the liquid jet. This phenomenon is important in many practical applications, for example, superheated liquid sprays in gasoline direct injection engines as well as cryogenic engines. The simulation of a flash boiling spray of such highly volatile liquid fuels is numerically challenging due to several reasons, including (1) the complexity of the bubble growth process in the presence of multiple vapor bubbles and (2) the need to use an extremely small time step size to accurately capture the underlying physics associated with the flash boiling process. In this paper, we first present a bubble growth model in flash boiling microdroplets considering bubble-bubble interactions along with the finite droplet size effects. A dimensional analysis of the newly derived Rayleigh-Plesset equation (RPE) with bubble-bubble interactions is then performed for Reynolds numbers of different orders of magnitude to estimate the relative importance of different forces acting on the bubble surface. Based on this analysis, a simplified nondimensional semi-analytical solution for bubble growth, which also includes the bubble-bubble interactions, is derived to estimate the bubble growth behavior with reasonable accuracy using the larger time step sizes for a wide range of operating conditions. The derived semi-analytical solution is shown to be a good approximation for describing the bubble growth rate  over the whole lifetime of the bubble, thus making it useful for simulations of superheated sprays with large numbers of droplets and even more bubbles. The bubble-bubble interactions are found to have a significant impact on the bubble growth dynamics and result in delaying the onset of droplet bursting due to the slower growth of the vapor bubble compared to the bubble growth without bubble-bubble interactions. Furthermore, in a comparison with DNS results, the proposed bubble growth model is shown to reasonably capture the impact of bubble interactions leading to smaller volumetric droplet expansion.
\end{abstract}


\end{frontmatter}


\section{Introduction}\label{intro}
The growth of a vapor bubble in a liquid is of great interest in the study of superheated boiling phenomena, e.g., flash boiling in gasoline direct injection engines and cryogenic engines. Generally, a bubble undergoes four different growth stages, i.e., (1) surface tension-controlled (ST) stage, (2) transition (T) stage, (3) inertia-controlled (IC) stage, and (4) thermal diffusion-controlled (TD) stage \citep{Robinson2002}. The bubble grows very slowly in the initial growth stage, which is mainly dominated by the surface tension force. A significant growth starts in the later ST-period due to strong positive thermal feedback from the surrounding superheated liquid. The transition growth stage is limited by both surface tension and inertia forces. As the bubble grows sufficiently large, in the IC-stage, the surface tension force diminishes, and only the inertial force limits the growth rate. Once the inertial forces tend towards zero, in the TD-stage, only the thermal diffusion acts as a rate-limiting factor for the bubble growth. It is to be noted that in cases, where the liquid inertial force is not sufficient to balance the pressure difference at the bubble surface, thermal diffusion along with the surface tension control the growth in the transition stage \citep{Robinson2002}, which thus connects the surface tension and thermal diffusion-controlled growth stages. \par
Several numerical studies are available in the literature, which simulate all the growth stages that a vapor bubble undergoes during its lifetime, such as \cite{Board1971}, \cite{Donne1975}, \cite{Lee1996}, and \cite{Robinson2002} to name a few. \cite{Board1971} developed a simple theory of bubble growth assuming temperature equilibrium at the liquid-vapor interface and a linear temperature drop across a thin thermal boundary layer surrounding the bubble surface. \cite{Donne1975}, \cite{Lee1996}, and \cite{Robinson2002} proposed similar bubble growth models by relaxing the assumption of the linear temperature drop across the thin thermal boundary layer. They numerically solved a one-dimensional energy equation together with the one-dimensional momentum equation such that the temperature gradient at the liquid-vapor interface can be obtained from the computed temperature field. The bubble evolutions predicted by their models were in good agreement with the experimental measurements. \par
The numerical studies described above were performed to simulate the growth of a single vapor bubble in a homogeneous infinite liquid medium. The liquid expansion and disruption phenomena were not considered in these studies. However, for flash boiling spray in automotive or cryogenic engines, multiple bubbles can be present inside superheated liquid droplets. As the vapor bubbles grow, the droplets continue to expand and can eventually rupture into smaller child droplets once a critical void fraction is reached. To explore the effect of the presence of multiple bubbles and the droplet disruption phenomena, \cite{Xi2017} simulated the flash boiling in a superheated single droplet of dimethyl ether ($\mathrm{DME}$) using the boiling explosion model derived by~\cite{Zhang2009}. They reported that the boiling explosion time becomes shorter for higher fuel temperatures and lower ambient pressures. \cite{Saha2022} recently proposed a reduced-order model (ROM) for simulating the flash boiling of a superheated single droplet of highly volatile e-fuels with multiple monodisperse vapor bubbles under subatmospheric operating conditions. They also investigated the detailed underlying physics associated with the growth of vapor bubbles. \par
The effect of bubble-bubble interactions is crucial in determining the growth characteristics of vapor bubbles in superheated microdroplets. \cite{Dietzel2019a} performed a direct numerical simulation (DNS) of bubble clusters to quantify the effects of bubble-bubble interactions in comparison with the single bubble growth for liquid oxygen (LOx) in different simplified configurations. They reported a significant reduction (more than 70\%) in the growth rate of the vapor bubbles located near the center of the LOx jet. These bubbles were found to have a negligible contribution to the volumetric expansion of the liquid jet. However, such fully-resolved simulations for accurate predictions of the bubble growth characteristics in multibubble environments are not feasible for large-scale spray simulations due to the high computational cost. A ROM considering the bubble-bubble interactions is thus necessary in this regard. The studies available in the literature mainly focus on the reduced-order modeling of the bubble-bubble interactions in the context of cavitation phenomena (such as \cite{Mettin1997}, \cite{Delale2001}, \cite{Bremond2006}, \cite{Ida2007}, \cite{Yi2016}, \cite{Maiga2018}, and \cite{Shi2021}, to name a few). In contrast, studies on reduced-order modeling of flash boiling of single droplets are still scarce in the literature. Although there are plenty of studies on flash boiling ROMs of single droplets and spray available in the literature (such as \cite{Senda1994}, \cite{Adachi1997}, \cite{Zuo2000}, \cite{Robinson2002}, \cite{Park2011}, \cite{Ji2011}, \cite{Yang2017}, \cite{Price2018}, \cite{Yu2020}, and \cite{Saha2022}, to name a few), none of them considered the effect of bubble-bubble interactions in their study.\par
The time-step size is another important factor in the simulation of the flash boiling phenomena of high-volatility liquid fuels. In a previous study by the current authors~\citep{Saha2022}, a fixed time-step size of one nanosecond $\left(\mathcal{O}\left(10^{-9}\,\text{s}\right)\right)$ was used to simulate the flash vaporization phenomena of highly volatile e-fuel microdroplets. Yet, it was found that this time-step size is still insufficiently small when considering bubble-bubble interactions for high-volatility fuels. Thus, the computational cost will also increase for the simulation of the large-scale spray case consisting of millions of single droplets.\par
A universal correlation or a simplified analytical solution for the bubble growth rate throughout the whole bubble lifetime, which also considers all the complex physical processes including bubble-bubble interactions occurring during the flash vaporization of a superheated single droplet, would be useful for the accurate and computationally efficient simulation of the large-scale spray case.\par
Previous studies were mainly focused on the derivation of the analytical formulation and correlation for the single vapor-bubble growth in an infinite superheated liquid medium only at later growth stages, such as IC and TD growth \citep{Strutt2009, Plesset1954, Scriven1959, Lien1969, Mikic1970, Prosperetti1978, Riznic1999}. \cite{Strutt2009} proposed an analytical solution for bubble growth by neglecting the viscous effect and the cooling effect of vaporization as 
\begin{equation}\label{eqn:rayleigh}
\frac{\mathrm d\Rb}{\mathrm dt}=\left(\frac{2}{3}\frac{\Pv-\Pl}{\rhol}\right)^{\frac{1}{2}},
\end{equation}
where $\Rb$ is the bubble radius, $t$ the time, and $\Pl$ the liquid pressure. $\rhol$ and $\Pv$ are the saturated liquid density and vapor pressure, respectively, at bulk liquid temperature $\Tl$. \refe{eqn:rayleigh} is referred to as inertia-controlled growth and provides a good description of bubble dynamics for sufficiently large initial superheating degrees or sufficiently low system pressures. \cite{Plesset1954} later obtained an asymptotic solution of bubble growth controlled by the thermal diffusion effect as
\begin{equation}\label{eqn:pless}
\frac{\mathrm d\Rb}{\mathrm dt}=\frac{1}{2}\left(\frac{12\alphal}{\pi t}\right)^\frac{1}{2}\frac{\rhol \cl\left(\Tl-\Tsat\left(\Pl\right)\right)}{\Lv \rhov}, 
\end{equation}
where $\alphal$ is the thermal diffusivity of the liquid and $\cl$ the liquid specific heat capacity. $\Lv$ and $\rhov$ denote the latent heat and saturated vapor density at saturation temperature, $\Tsat$, respectively. The time $t$ in \refe{eqn:pless} is measured from the onset of nucleation of the bubble. The range of applicability of the analytical works described above is limited to the later stages of bubble growth when the radius is much larger than the critical radius defined as
\begin{equation}\label{Rcrit}
\Rc=\frac{2\sigma\left(\Tl\right)}{\Pv -\Pl},
\end{equation}
where $\sigma$ is the surface tension.\par
\par
In order to correlate the bubble radius as a function of time in terms of nondimensional parameters, \cite{Lien1969} derived the reference length and time scales by requiring the nondimensionalization to be reasonable for both \refe{eqn:rayleigh} and \refe{eqn:pless}, which leads to
\begin{equation}\label{eqn:ref_scales}
\Rref = \frac{B^{2}}{A}, \quad \tref=\frac{B^{2}}{A^{2}},
\end{equation}
with 
\begin{equation}\label{eqn:LienAB}
\begin{split}
A&=\left({\frac{2}{3} \frac{\Pv-\Pl}{\rhol}}\right)^{\frac{1}{2}},  \quad B=Ja \left({\frac{12}{\pi} \alphal}\right)^{\frac{1}{2}},
\end{split}
\end{equation}
 where $Ja$ is the Jakob number defined as 
\begin{equation}
Ja=\frac{\rhol \cl\left(\Tl-\Tsat \left(\Pl \right)\right)}{\rhov \Lv}.
\end{equation}\par
The nondimensional radius and time are then expressed as
\begin{equation}\label{eqn:AB}
\Rbstar = \frac{\Rb}{\Rref} \quad \mathrm{and} \quad \tstar=\frac{t}{\tref}.
\end{equation}
\par
\cite{Mikic1970} reported an analytical formulation for bubble growth rate by combining \refe{eqn:rayleigh} and \refe{eqn:pless} in terms of nondimensional parameters proposed by \cite{Lien1969} as
\begin{equation}
    \frac{\text d\Rbstar}{\text dt^*}=\sqrt{t^*+1}-\sqrt{t^*}.
\end{equation} 
\par
The scaling proposed by \cite{Lien1969} works well only in the later period of the bubble lifetime when the growth is mainly controlled by inertia or thermal diffusion effect. \cite{Prosperetti1978} later attempted to derive an approximate law of bubble growth based on the theory provided by \cite{Plesset1954} for the bubbles that have grown sufficiently large beyond their critical radius, at least by about an order of magnitude. In addition, they also defined a more general scaling law, which was expected to work also in the surface tension-dominated growth phase, However, the detailed analysis of this scaling law was not reported as they found it to be of less practical interest compared to the former one. The scaling relationships proposed by \cite{Prosperetti1978} are given in \ref{appA}.\par 
The surface tension-controlled growth may become relevant to automotive or cryogenic engine applications, such as injecting superheated liquid fuel into a combustion chamber. Here, the lifetime of the vapor bubbles becomes much shorter as the liquid jet undergoes bursting phenomena once the critical void fraction is reached~\citep{Senda1994}. \cite{Saha2022} reported that for moderate to high superheating degrees, the onset of droplet burst times, $t_\mathrm b$, for highly volatile e-fuel microdroplets are in the order of $\mathcal{O}(t_\mathrm b)\approx10^{-6}-10^{-7}$ s. For the lower superheating degree case, the order of magnitude increases to $\mathcal{O}(t_\mathrm b)\approx10^{-4}$ s. Due to the shorter droplet lifetime, the initial surface tension-controlled growth phase of the vapor bubbles becomes equally important to the inertia, transition, and thermal diffusion-controlled growth phases. This is because the shorter the ST growth stage is, the earlier the significant bubble growth starts and vice-versa, thus influencing the bubble-growth characteristics in the later bubble growth stages as well. However, the influence of the ST growth stage on the bubble dynamics becomes negligible when the bubble grows for a long period \citep{Lee1996}.\par
Recently, \cite{Saha2021} derived two different analytical solutions for the vapor bubble growth rate in superheated single droplets for high and low Reynolds numbers (Re) using a similar general scaling law to that proposed by \cite{Prosperetti1978}, but neglected the bubble-bubble interactions in their study. The analytical solutions derived by \cite{Saha2021} are provided in \ref{appA}.\par
Along these lines, the objectives of this paper can be summarized as follows: (1) to propose a modified Rayleigh-Plesset equation (RPE) for characterizing the growth of vapor bubbles in superheated microdroplets considering bubble-bubble interactions; (2) to highlight the limitations of the numerical solver for simulating the bubble growth with the newly derived RPE in a flash boiling single droplet of highly volatile liquid fuels; (3) to analyze the effect of bubble-bubble interactions on the growth characteristics of the vapor bubbles in superheated microdroplets; (4) to explore a scaling law for evaluating its capability of describing the different bubble growth stages including ST, T, IC, and TD growth for a flash boiling single droplet; and (5) to derive a semi-analytical solution in terms of nondimensional parameters for the bubble growth considering the bubble-bubble interactions to describe the bubble evolution with reasonable accuracy in different growth phases. \par
The manuscript is organized as follows: in Section~\ref{ThetForm}, first, we present the bubble growth models for superheated microdroplets with and without considering bubble-bubble interactions. Next, we provide the nondimensional formulations of the governing equations. The details of the computational setup and the solution procedure are shown in Section~\ref{CompStSolPrc}. The performance of the proposed bubble growth model is compared with that of the DNS results by \cite{Dietzel2019a} in Section~\ref{Validation}. Section~\ref{Reslt} outlines the results from the numerical simulations. Finally, the findings are summarized in Section~\ref{Conc}.

\section{Theoretical formulations} \addvspace{10pt}
\label{ThetForm}
The conventional Rayleigh-Plesset equation (RPE) \citep{Brennen2013} is often used to describe the temporal dynamics of a single vapor bubble in a superheated liquid medium. However, the conventional RPE needs to be modified for the flash boiling microdroplets containing multiple bubbles, where the interaction among bubbles significantly alters the bubble dynamics~\citep{Saha2022}. In this study, we first derive the modified RPE considering the bubble-bubble interactions along with the finite-size effects of the flash boiling microdroplets. \par
In this section, first, we discuss the governing equations of the single isolated vapor bubble growth in an infinite liquid medium and in a finite-size superheated microdroplet. Next, we describe the equations governing the vapor bubble growth in superheated microdroplets considering the bubble-bubble interactions. Finally, we present the nondimensional form of the governing equations.\par
\subsection{Single isolated vapor bubble growth}\addvspace{10pt}
The growth rate of an isolated spherically symmetric vapor bubble in a homogeneous infinite liquid medium can be estimated using the conventional RPE as~\citep{Brennen2013}
\begin{equation}\label{RPE1}
\begin{split}
\underbrace{\Pv\left(\Tv\right)-\Pl}_{\DelP} & = \underbrace{\rhol\Rb \frac{\mathrm{d}^{2} \Rb}{\mathrm{~d} t^{2}}}_{\Phdaccb}+\underbrace{\frac{3}{2}\rhol\left(\frac{\mathrm{d} \Rb}{\mathrm{~d} t}\right)^{2}}_{\Phdvelb}+\underbrace{\frac{4{\mul}}{\Rb} \frac{\mathrm{d} \Rb}{\mathrm{d} t}}_{\Pmu}+\underbrace{\frac{2 \sigma}{\Rb}}_{\Psigma},
\end{split}
\end{equation}
 where $\Rb (t)$ denotes the instantaneous vapor bubble radius and $\mul$ the liquid viscosity. The physical parameters of \refe{RPE1} are evaluated at $\Tv$, which is assumed to be equal to the liquid temperature at the liquid-vapor interface~\citep{Saha2022}. In \refe{RPE1}, $\DelP$ describes the difference between saturated vapor pressure inside the vapor bubble and liquid pressure acting on the bubble surface, $\Pmu$ the pressure due to liquid viscosity, and $\Psigma$ the pressure due to the surface tension force at the liquid-vapor interface. $\Phdaccb$ and $\Phdvelb$ are the hydrodynamic pressures induced by the surrounding liquid due to the bubble surface acceleration and velocity, respectively. In this study, we denote the total hydrodynamic pressure for the bubble growth without bubble-bubble interactions as $\Phd=\Phdaccb+\Phdvelb$.\par
For a finite-size single droplet in a quiescent gaseous medium, the conventional RPE can be modified as
\begin{equation}\label{RPEtrl}
   \begin{split}
       \Pv\left(\Tv\right)-\Pl=\rhol\left[\Rb\left(1-\frac{\Rb}{\Rd}\right) \frac{\mathrm{d}^{2}\Rb}{\mathrm{~d} t^{2}}+\left(\frac{3}{2}-\frac{2 \Rb}{\Rd}+\frac{\Rb^{4}}{2 \Rd^{4}}\right)\left(\frac{\mathrm{d} \Rb}{\mathrm{~d} t}\right)^{2}\right]+\frac{4 \mu_{l}}{\Rb} \frac{\mathrm{d} \Rb}{\mathrm{~d} t}+\frac{2\sigma}{\Rb},
   \end{split}
\end{equation}
where $\Rd (t)$ denotes the instantaneous droplet radius. In comparison with the conventional RPE shown in \refe{RPE1}, \refe{RPEtrl} includes additional inertia terms on the right-hand side given by $-\rhol\left(\Rb^2/\Rd\right)\left(\mathrm{d^2}\Rb/\mathrm{d}t^2\right)$ and $\rhol\left\{-2\Rb/\Rd+\Rb^4/\left(2\Rd^4\right)\right\}\left(\mathrm{d}\Rb/\mathrm{d}t\right)^2$. $\Pl$ in \refe{RPEtrl} is given by the mechanical balance between the liquid and the surrounding gas phase as
\begin{equation}
    \Pl=\Pg+\frac{2 \sigma}{\Rd},
\end{equation}
where $\Pg$ is the surrounding gas pressure. Substituting $\Pl$ into \refe{RPEtrl} yields
\begin{equation}\label{RPE2}
   \begin{split}
       \underbrace{\Pv\left(\Tv\right)-\Pg}_{\DelP}=\underbrace{\rhol\Rb\left(1-\frac{\Rb}{\Rd}\right) \frac{\mathrm{d}^{2} \Rb}{\mathrm{~d} t^{2}}}_{\Phdaccb}+\underbrace{\rhol\left(\frac{3}{2}-\frac{2 \Rb}{\Rd}+\frac{\Rb^{4}}{2 \Rd^{4}}\right)\left(\frac{\mathrm{d} \Rb}{\mathrm{~d} t}\right)^{2}}_{\Phdvelb}+\underbrace{\frac{4 \mu_{l}}{\Rb} \frac{\mathrm{d} \Rb}{\mathrm{~d} t}}_{\Pmu}+\underbrace{2 \sigma\left(\frac{1}{\Rb}+\frac{1}{\Rd}\right)}_{\Psigma}.
   \end{split}
\end{equation}
It is to be noted that the \refe{RPE2} is valid only for a single isolated vapor bubble inside a superheated single droplet.
\subsection{Vapor bubble growth in a multibubble environment}
\begin{figure}[!h]
\centering
\includegraphics[width=180pt]{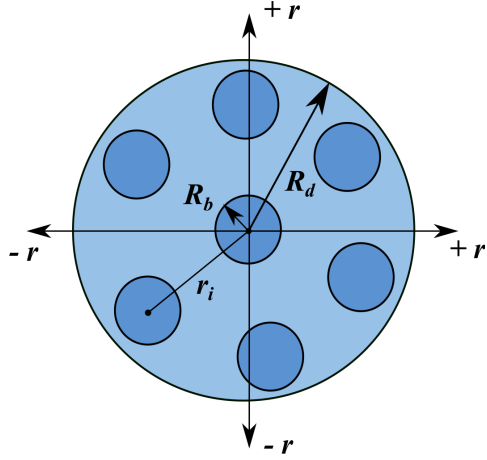}
\caption{Schematic of the reduced-order modeling of bubble-bubble interactions in superheated microdroplets.}
\vspace{-0.08 in}
\label{Schm_Int}
\end{figure}

\addvspace{10pt}
In reality, there are multiple bubbles inside the droplet. Thus, the bubble growth rate is also influenced by the interaction between the individual bubbles. Considering the bubble-bubble interactions, the RPE in \refe{RPE2} is modified as
\begin{equation}\label{RPE3}
\begin{split}
\Pv\left(\Tv\right)-\Pg=\rhol\left[\frac{\mathrm{d}}{\mathrm{d} t}\left(\sum \frac{\Rb^{2}}{\Ri} \frac{\mathrm{d} \Rb}{\mathrm{~d} t} \right)+\Rb\left(1-\frac{\Rb}{\Rd}\right) \frac{\mathrm{d}^{2} \Rb}{\mathrm{~d} t^{2}}+\left(\frac{3}{2}-\frac{2 \Rb}{\Rd}+\frac{\Rb^{4}}{2 \Rd^{4}}\right)\left(\frac{\mathrm{d} \Rb}{\mathrm{~d} t}\right)^{2}\right]
+\frac{4 \mul}{\Rb} \frac{\mathrm{d} \Rb}{\mathrm{~d} t}\\+2 \sigma\left(\frac{1}{\Rb}+\frac{1}{\Rd}\right),
\end{split}
\end{equation}
where the first term on the right-hand side of \refe{RPE3} represents the pressure, $\Pinertia$, acting on the target bubble surface induced by all the other bubbles. The distance $\Ri$ is measured from the center of the $i$-th bubble to the center of the target  bubble. The bubbles are assumed to be monodisperse in this study. For simplicity, it is also assumed that the bubbles remain sufficiently far away from each other such that the interactions among the boundary layers across the bubble surface can be neglected~\citep{Saha2022} and the relative position of the bubbles remain unchanged during the vaporization process. A schematic model of the bubble-bubble interactions is shown in \reff{Schm_Int}. 
The detailed derivation of $\Pinertia$ is given in \ref{appB}. $\Pinertia$ can be rearranged as follows~\citep{Kubota1992}
\begin{align}\label{prss1}
    \frac{\mathrm{d}}{\mathrm{d} t}\left[\sum \frac{\Rb^{2}}{\Ri} \frac{\mathrm{d} \Rb}{\mathrm{~d} t} \right]=\frac{\mathrm{d}}{\mathrm{d} t}\left[\Rb^{2}\frac{\mathrm{d} \Rb}{\mathrm{~d} t}  \sum \frac{1}{\Ri}\right]=\frac{\mathrm{d}}{\mathrm{d} t}\left[\Rb^{2}\frac{\mathrm{d} \Rb}{\mathrm{~d} t}  \int_{0}^{\Rd} \frac{1}{r}(n \mathrm{~d} v)\right],
\end{align}
where $n$ denotes the number density of the vapor bubbles and $n\mathrm{~d} v$ the total number of bubbles in an infinitesimal volume of the droplet, $\mathrm{~d}v$.\par
Assuming heterogeneous nucleation due to air as dissolved gas, the number density of the vapor bubble nuclei can be estimated using the approximation suggested by \cite{Senda1994} as
\begin{equation}\label{NumDen}
n=5.757 \times 10^{12} \cdot \exp \left(\frac{-5.279 \mathrm{~K}}{\Delta \theta}\right),
\end{equation}
where $\Delta\theta$ denotes the superheating degree defined as $\Delta\theta$ = $\Td-\Tb(\Pg)$, with $\Td$ the droplet bulk temperature and $\Tb$ the droplet boiling temperature at $\Pg$.\par
The integration over the sphere of influence ($\Rd$) in \refe{prss1} yields
\begin{align}\label{prss2}
    \begin{aligned} \frac{\mathrm{d}}{\mathrm{d} t}\left[\Rb^{2}\frac{\mathrm{d} \Rb}{\mathrm{~d} t}  \int_{0}^{\Rd} \frac{1}{r}(n \mathrm{~d} v)\right] &=\frac{\mathrm{d}}{\mathrm{d} t}\left[\Rb^{2}n\frac{\mathrm{d} \Rb}{\mathrm{~d} t} \int_{0}^{\Rd} \frac{1}{r} 4 \pi r^{2} \mathrm{~d} r\right] \\ &=\frac{\mathrm{d}}{\mathrm{d} t}\left[2\pi\Rb^{2} n\frac{\mathrm{d} \Rb}{\mathrm{~d} t}\Rd^{2}\right] \\ &=2 \pi n \Rd\left[\Rd \Rb\left\{\Rb \frac{\mathrm{d}^{2} \Rb}{\mathrm{~d} t^{2}}+2\left(\frac{\mathrm{d} \Rb}{\mathrm{~d} t}\right)^{2}\right\}+2 \Rb^{2}\left(\frac{\mathrm{d} \Rb}{\mathrm{~d} t}\right)\left(\frac{\mathrm{d} \Rd}{\mathrm{~d} t}\right)\right], \end{aligned}
\end{align}
where it is assumed that the number of vapor bubbles in the droplet remains constant during the flash vaporization process. Substituting \refe{prss2} into \refe{RPE3} and rearranging the terms, the modified RPE for spherically symmetric vapor bubbles in a single droplet of radius $\Rd$ considering the bubble-bubble interactions is written as
\begin{equation}\label{RPE4}
\begin{split}
    \underbrace{\Pv\left(\Tv\right)-\Pg}_{\DelPprime}=\underbrace{\rhol\Rb\left(1-\frac{\Rb}{\Rd}+2 \pi \Rd^{2} n \Rb\right) \frac{\mathrm{d}^{2} \Rb}{\mathrm{~d} t^{2}} }_{\Phdaccbprime}+\underbrace{\rhol\left(\frac{3}{2}-\frac{2 \Rb}{\Rd}+\frac{\Rb^{4}}{2 \Rd^{4}}+4 \pi \Rd^{2} n \Rb\right)\left(\frac{\mathrm{d} \Rb}{\mathrm{~d} t}\right)^{2}}_{\Phdvelbprime}+\underbrace{\frac{4\mul}{\Rb}\frac{\mathrm{d} \Rb}{\mathrm{~d} t}}_{\Pmuprime}\\+\underbrace{4 \pi n \rhol \Rd \Rb^{2} \left(\frac{\mathrm{d} \Rd}{\mathrm{~d} t}\right)\left(\frac{\mathrm{d} \Rb}{\mathrm{~d} t}\right)}_{\Phdveldbprime}+\underbrace{2 \sigma\left(\frac{1}{\Rb}+\frac{1}{\Rd}\right)}_{\Psigmaprime},
\end{split}
\end{equation}
where the superscript ` $\prime$ ' denotes the modified pressures acting on the bubble surface for the bubble growth considering the bubble-bubble interactions. The new pressure term, $\Phdveldbprime$, is the contribution to the total hydrodynamic pressure ($\Phdprime$) due to the expansion velocities of both the bubble and the droplet. $\Phdprime$ is thus expressed as $\Phdprime=\Phdaccbprime+\Phdvelbprime+\Phdveldbprime$.\par
In this study, the droplet is considered much larger than the vapor bubbles ($\Rd\gg\Rb$). Thus, the second term of $\Phdaccbprime$ can be neglected, since the order of magnitude of the bubble to droplet radius ratio, $\Rb/\Rd$,  remains very small compared to the first and third terms. Similarly, the contribution from the second and third terms of $\Phdvelbprime$ also becomes negligible with respect to the first and fourth terms and hence can be neglected. The second term of $\Psigmaprime$ can also be neglected, since $1/\Rb\gg1/\Rd$. The resulting simplified RPE after neglecting the above-mentioned terms is given as 
\begin{equation}\label{RPE5}
\begin{split}
    \underbrace{\Pv\left(\Tv\right)-\Pg}_{\DelPprime}=\underbrace{\rho_{l}\Rb\left(1+2 \pi \Rd^{2} n \Rb\right) \frac{\mathrm{d}^{2} \Rb}{\mathrm{~d} t^{2}}}_{\Phdaccbprime}+\underbrace{\rhol\left(\frac{3}{2}+4 \pi \Rd^{2} n \Rb\right)\left(\frac{\mathrm{d} \Rb}{\mathrm{~d} t}\right)^{2}}_{\Phdvelbprime}+\underbrace{\frac{4 \mul}{\Rb}\frac{\mathrm{d} \Rb}{\mathrm{~d} t}}_{\Pmuprime}\\+\underbrace{4 \pi n \rhol \Rd \Rb^{2} \left(\frac{\mathrm{d} \Rd}{\mathrm{~d} t}\right) \left(\frac{\mathrm{d} \Rb}{\mathrm{~d} t}\right)}_{\Phdveldbprime}+\underbrace{\frac{2\sigma}{\Rb}}_{\Psigmaprime}.
\end{split}
\end{equation}
The saturated vapor temperature, $\Tv$, is obtained by solving the implicit relation as~\citep{Saha2022}
\begin{equation}\label{eqn:vap_tmp}
\frac{\Pv W}{\Rgas \Tv} \Lv \frac{\mathrm{d} \Rb}{\mathrm{d} t}+\Lv \frac{\Rb}{3} \frac{\mathrm{d}}{\mathrm{d} t}\left(\frac{\Pv W}{\Rgas \Tv}\right)=\lambdal \frac{\Td-\Tv}{\delta},
\end{equation}
where $W$ is the molecular weight of the fluid, $\lambdal$ the liquid thermal conductivity, and $\Rgas$ the universal gas constant. $\delta$ in \refe{eqn:vap_tmp} denotes the thickness of the thin thermal boundary layer adjacent to the bubble surface defined as~\citep{Board1971} 
\begin{equation}
\delta=\left({\frac{\pi\alphal t}{3}}\right)^{\frac{1}{2}}.
\end{equation}
The energy flux at the liquid-vapor interface causes the liquid to vaporize and creates a pressure difference, which is then balanced in the RPE by the hydrodynamic terms (IC growth stage), the surface tension term (ST growth stage)), or both (T stage).\par
In this study, the vapor bubbles are assumed to grow until a critical void fraction value is reached. The void fraction, $\varepsilon$, is defined as~\citep{Senda1994}
\begin{equation}
\varepsilon=\frac{V_\mathrm{bubble}}{V_\mathrm{bubble}+V_\mathrm{droplet}},
\end{equation}
where $V_\mathrm{droplet}$ describes the volume of the liquid droplet and $V_\mathrm{bubble}$ the total volume of the vapor bubbles. The droplet is considered to burst once the void fraction exceeds its critical limit ($\varepsilon\ge\varepsilon_\mathrm{crit}$). The critical void fraction value of 0.55 is chosen here for all simulated test cases~\citep{Kawano2004}.
\subsection{Nondimensionalization of the governing equations} \addvspace{10pt}
The RPE is nondimensionalized using the critical bubble radius, $\Rc$,  as the length scale, $A$ as the velocity scale (\refe{eqn:LienAB}), and the time scale, $\tref=\Rc/A$~\citep{Saha2021}:
\begin{equation}\label{eqn:scl_qnt1}
\begin{split}
\Rplus_\text{k} &=\frac{R_\text{k}}{\Rc}, \quad \dot R_\text{k}^+ = \frac{\dot R_\text{k}}{A}, \quad \tplus=\frac{t}{\tref}, \quad \mulplus=\frac{\mul}{\mulzero};\\
\rholplus &=\frac{\rhol}{\rholzero}, \quad \Pplus =\frac{P}{\rholzero A^{2}}, \quad \sigmaplus =\frac{\sigma}{\sigma_0}, \quad \Ndenplus=n\Rc^3,
\end{split}
\end{equation}
where the subscript `k' can be replaced by subscripts `d' and `b' for droplet and bubble quantities, respectively. The superscript ‘+’ denotes the nondimensional quantities and the subscript ‘0’ represents the initial values.\par 
The resulting nondimensional RPE for a spherically symmetric vapor bubble, considering the bubble-bubble interactions, in a finite-size single droplet of radius $\Rd$ is given as 
\begin{equation}\label{RPEndwB}
   \begin{split}
     \underbrace{\Pvplus-\Pgplus}_{\DelPplusprime }=\underbrace{\rholplus\Rbplus\left(1+2 \pi \Rdplustwo \Ndenplus \Rbplus\right) \frac{\mathrm{d}^{2} \Rbplus}{\mathrm{~d}t^{+2}}}_{\Phdaccbplusprime}+\underbrace{\rholplus\left(\frac{3}{2}+4 \pi \Rdplustwo \Ndenplus \Rbplus\right)\left(\frac{\mathrm{d} \Rbplus}{\mathrm{~d} \tplus}\right)^{2}}_{\Phdvelbplusprime}+\underbrace{\frac{4}{\Rbplus} \frac{\mulplus}{\text{Re}}\frac{\mathrm{d} \Rbplus}{\mathrm{~d} \tplus }}_{\Pmuplusprime}\\+\underbrace{4 \pi \Ndenplus \rholplus \Rdplus \Rbplustwo \left(\frac{\mathrm{d} \Rdplus}{\mathrm{~d} \tplus}\right) \left(\frac{\mathrm{d} \Rbplus}{\mathrm{~d} \tplus}\right)}_{\Phdveldbplusprime}+\underbrace{\frac{2\sigmaplus}{\text{We}}\frac{1}{\Rbplus}}_{\Psigmaplusprime}.
  \end{split}
\end{equation}
\refe{RPEndwB} can be simplified for the bubble growth without bubble-bubble interactions as
\begin{equation}\label{eqn:nd_RPE1}
\begin{split}
\underbrace{\Pvplus-\Pgplus}_{\DelPplus}&=\underbrace{\rholplus\Rbplus \frac{\mathrm{d}^{2} \Rbplus}{\mathrm{d} \tplustwo}}_{\Phdaccbplus}+\underbrace{\rhol^+\frac{3}{2}\left(\frac{\mathrm{d} \Rbplus}{\mathrm{d} \tplus}\right)^{2}}_{\Phdvelbplus}+\underbrace{\frac{4}{\Rbplus}\frac{\mulplus}{\text{Re}}\frac{\mathrm{d} \Rbplus}{\mathrm{d} \tplus}}_{\Pmuplus}+\underbrace{\frac{2 \sigmaplus}{\text{We}}\frac{1}{\Rbplus}}_{\Psigmaplus}.
\end{split}
\end{equation}
Re and We are the Reynolds number and Weber number defined as
\begin{equation}\label{ReWe}
    \mathrm{Re}=\frac{\rho_{\mathrm{l}} A R_{\mathrm{c}}}{\mu_{\mathrm{l}}}\hskip0.5cm\text{and}\hskip0.15cm\quad \mathrm{We}=\frac{\rho_{\mathrm{l}} A^2 R_{\mathrm{c}}}{\sigma_0}=\frac{4}{3}.
\end{equation}
The derivation of the nondimensional RPE with and without bubble-bubble interactions is given in \ref{appC}.

\section{Numerical methodology}
\label{CompStSolPrc}
\begin{figure}[!t]
\centering
\includegraphics[width=310pt]{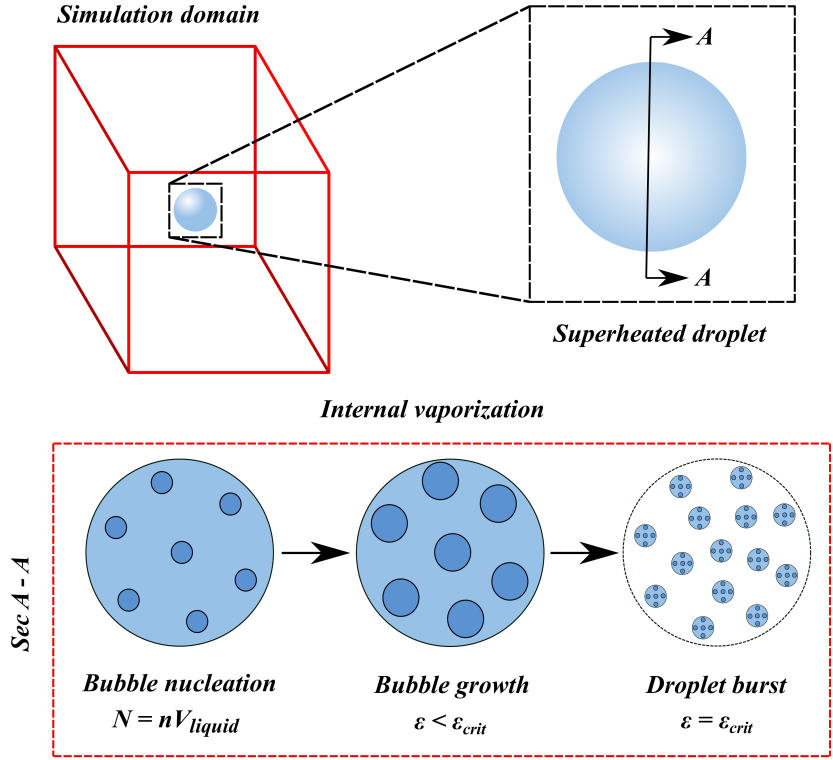}
\caption{Schematic of the simulation domain and single droplet flash boiling processes.}
\label{Schm}
\end{figure}

In this work, a square box of size 0.54 m $\times$ 0.54 m $\times$ 0.54 m with no-slip walls is chosen as the simulation domain. The schematic of the simulation domain is shown in \reff{Schm}. A superheated single droplet with a diameter of 200 $\upmu$m is placed at the center of the domain. Although a spherical droplet is introduced into a three-dimensional computational domain, the complete problem is reduced to the solution of an ordinary differential equation in zero dimension by assuming the spherical symmetry of the droplet and considering the surrounding flow field as frozen. 
The droplet is considered to be stationary. Inert gas nitrogen is used as an ambient medium and the chemistry in the gas phase is neglected. The temperature and pressure inside the liquid droplet are assumed to be homogeneous. The vaporization from the droplet's external surface is neglected in this study due to its negligible contribution compared to internal vaporization. The reader is referred to \cite{Saha2022} for a more detailed discussion regarding coupled internal and external vaporization under superheated operating conditions. It is to be noted that the size of the square box and the subsequent spatial discretization were chosen such that they have no influence on the bubble dynamics presented in this study due to the assumptions of the frozen surrounding flow field.\\
The nondimensional RPEs are advanced using a fixed time-step second-order Runge-Kutta (RK) scheme. \refe{RPEndwB} is solved for the case with bubble-bubble interactions. Whereas for the case without bubble-bubble interactions, \refe{eqn:nd_RPE1} is solved assuming monodisperse vapor bubbles of the number density of $n$ in an infinitely large single droplet ($\Rd\gg\Rb$). The bubble growth is initiated by perturbing the bubble critical radius, $\Rc$, by $\mathrm{10^{-4}\%}$. A detailed discussion on the influence of initial perturbation on the bubble dynamics has been provided by \cite{Lee1996} and \cite{Saha2022}. The temperature inside the vapor bubble, $\Tv$, is obtained by iteratively solving \refe{eqn:vap_tmp} using the Newton-Raphson method.

\section{Model validation}\label{Validation}
\begin{figure}[!t]
\centering
\includegraphics[width=420pt]{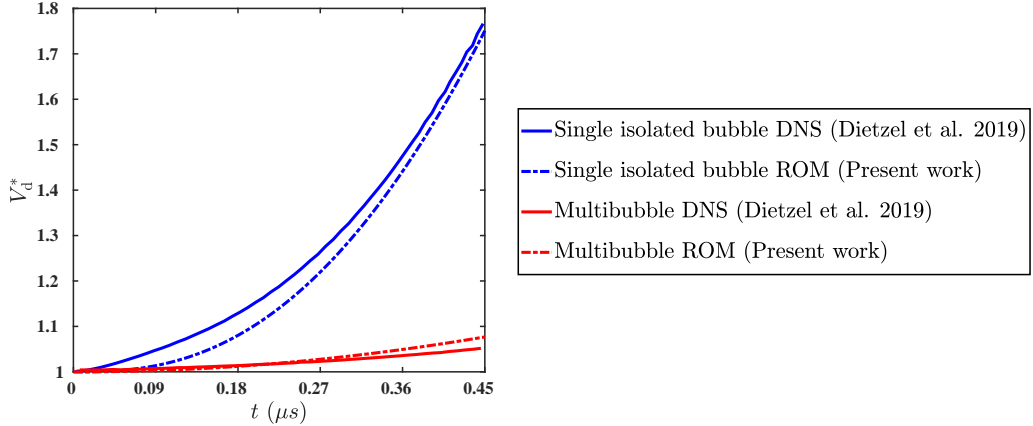}
\vspace{-0.3 in}
\caption{Comparison of the dimensionless volume of a LOx fuel droplet predicted by the proposed ROM with the DNS results reported by \cite{Dietzel2019a} for the bubble growth with (referred to as `multibubble') and without (referred to as `single isolated bubble') the bubble-bubble interactions at $\Pg=2.045$ bar and $\Td=120$ K. The initial volume of the liquid droplet is used as a reference scale to nondimensionalize the instantaneous droplet volume, $V_\text{droplet}$.}
\label{DNSVald2}
\end{figure}

Ideally, the proposed ROM needs to be validated at a microscopic level in two steps. First, the validation with experimental measurements of single isolated vapor bubbles is required to ensure that the model is able to accurately capture the single bubble growth in a superheated liquid medium. In the second step, the validation must be performed using flash boiling single droplet experimental measurements in the presence of multiple bubbles. The present model has already been validated in a previous publication~\citep{Saha2022} at the single isolated bubble level, which is shown here for completeness in \ref{appD}, and a further comparison with the single isolated bubble simulation by \cite{Dietzel2019a} will be shown in the following. Additionally, the performance of the proposed reduced-order bubble growth model with bubble-bubble interactions is compared against the DNS of multiple bubbles with variable expansion rate performed by \cite{Dietzel2019a} in terms of the volumetric expansion of a superheated LOx fuel droplet. A single droplet including 221 vapor bubbles was simulated for this purpose. The initial perturbation in $\Rc$ was chosen based on their single isolated bubble reference solution. The volume of the droplet was considered to be equivalent to that of the cylindrical liquid jet of \cite{Dietzel2019a}. \reff{DNSVald2} depicts the comparison of the nondimensional volumetric expansion of the LOx droplet for the isolated bubble and the bubble-bubble interaction case with the DNS results. For the single isolated bubble simulations, both the DNS and the ROM predict a similar volumetric expansion of around 75 \%, as can be seen from \reff{DNSVald2}. It is observed that the present ROM predicts a volumetric droplet expansion of around 7.5 \% for the case with bubble-bubble interactions, whereas the multibubble DNS with variable expansion rate results in approximately a 5 \% increase in the volume of the liquid jet. This demonstrates the strong impact of bubble interactions in the present case. The simulation is continued only up to $t=0.45\,\upmu$s, as this marks the minimum time required to the onset of bubble merging for the \cite{Dietzel2019a} case. The resulting overprediction from the ROM in the volumetric expansion for the bubble-bubble interactions case is reasonably small compared with the total effect of bubble interactions and can be attributed to the assumptions considered in its derivation such as spherically symmetric monodisperse vapor bubbles and homogeneous pressure and temperature distribution in the liquid droplet. Overall, it can be concluded that the proposed ROM is capable of predicting the considerably smaller volume expansion in presence of the bubble-bubble interactions with reasonable accuracy. \par 
At a macroscopic scale, the present bubble growth model could be validated by applying it to the primary breakup simulation of the flash boiling spray case and comparing the droplet size distributions in the near-nozzle regime obtained from the simulations with the experiments. This validation at a macroscopic level is beyond the scope of the present article and will be left for future work.
\section{Results and discussion} \addvspace{10pt}
\label{Reslt}
In order to accurately and efficiently describe the vapor bubble growth rate in superheated microdroplets, it is necessary to consider the following three sub-problems: (1) the limitations of the numerical solver, (2) the influence of bubble-bubble interactions, and (3) nondimensional analysis of vapor bubble growth. In this section, considering high volatility e-fuels, such as short-chain oxymethylene ethers ($\mathrm{OME_x}$) as a generic example, first, the issues related to the use of large time-steps are illustrated for the simulation of highly volatile liquid fuels under various operating conditions considering the bubble-bubble interactions. Next, the influence of the bubble-bubble interactions on the dynamics of the vapor bubble growth in superheated microdroplets is described. Then, the applicability of different scaling laws in describing the bubble growth in a flash boiling single droplet is highlighted under subatmospheric operating conditions. After that, the nondimensional RPE is used to evaluate the relative importance of different forces acting on the bubble surface. Finally, based on the nondimensional analysis, a semi-analytical solution is proposed for the bubble growth in terms of nondimensional parameters and validated with the numerical solution of the RPE.
\subsection{Limitations of the numerical solver}\label{Limit}
\begin{figure}[!t]
\centering
\includegraphics[width=420pt]{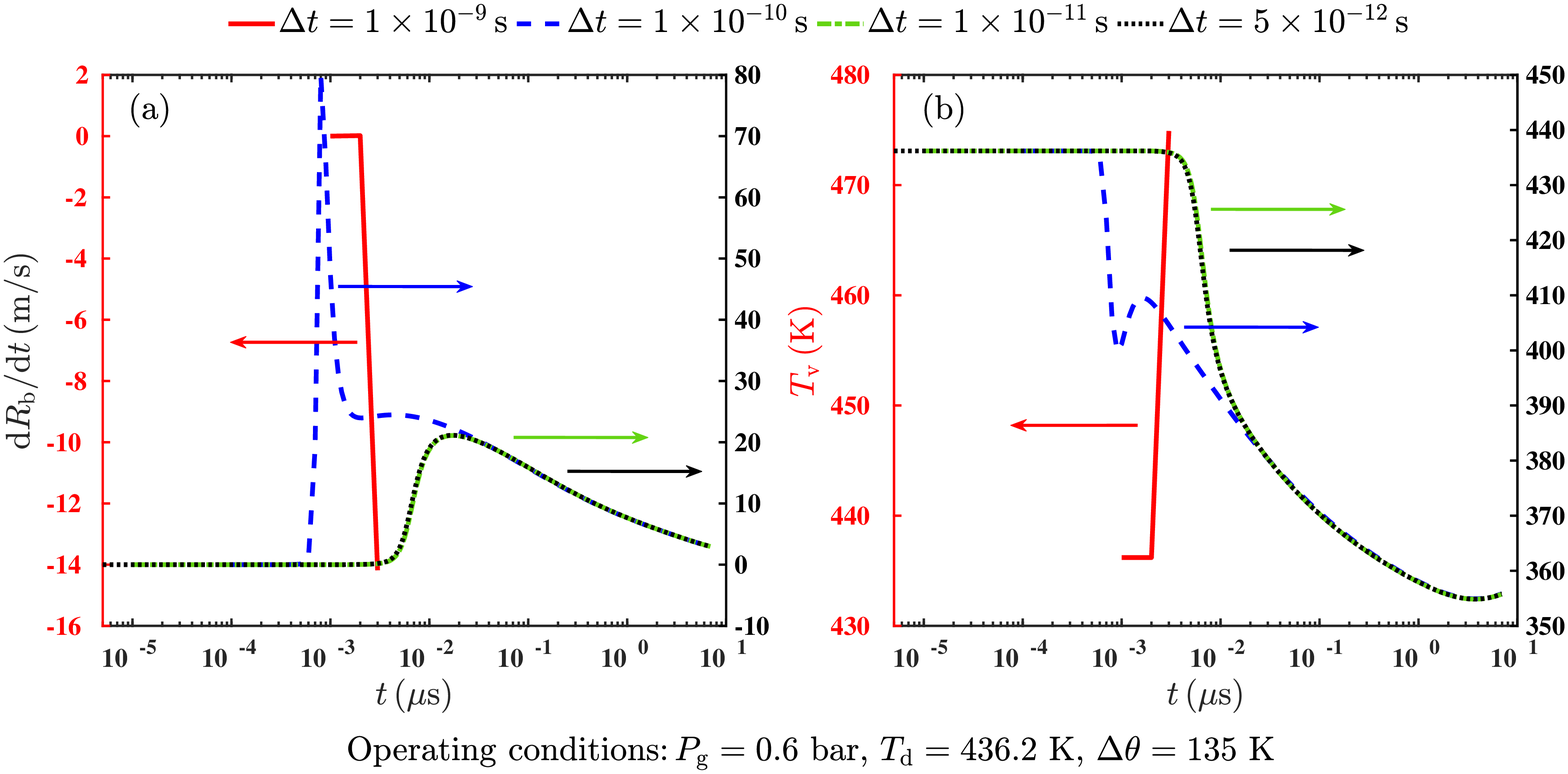} 
\includegraphics[width=420pt]{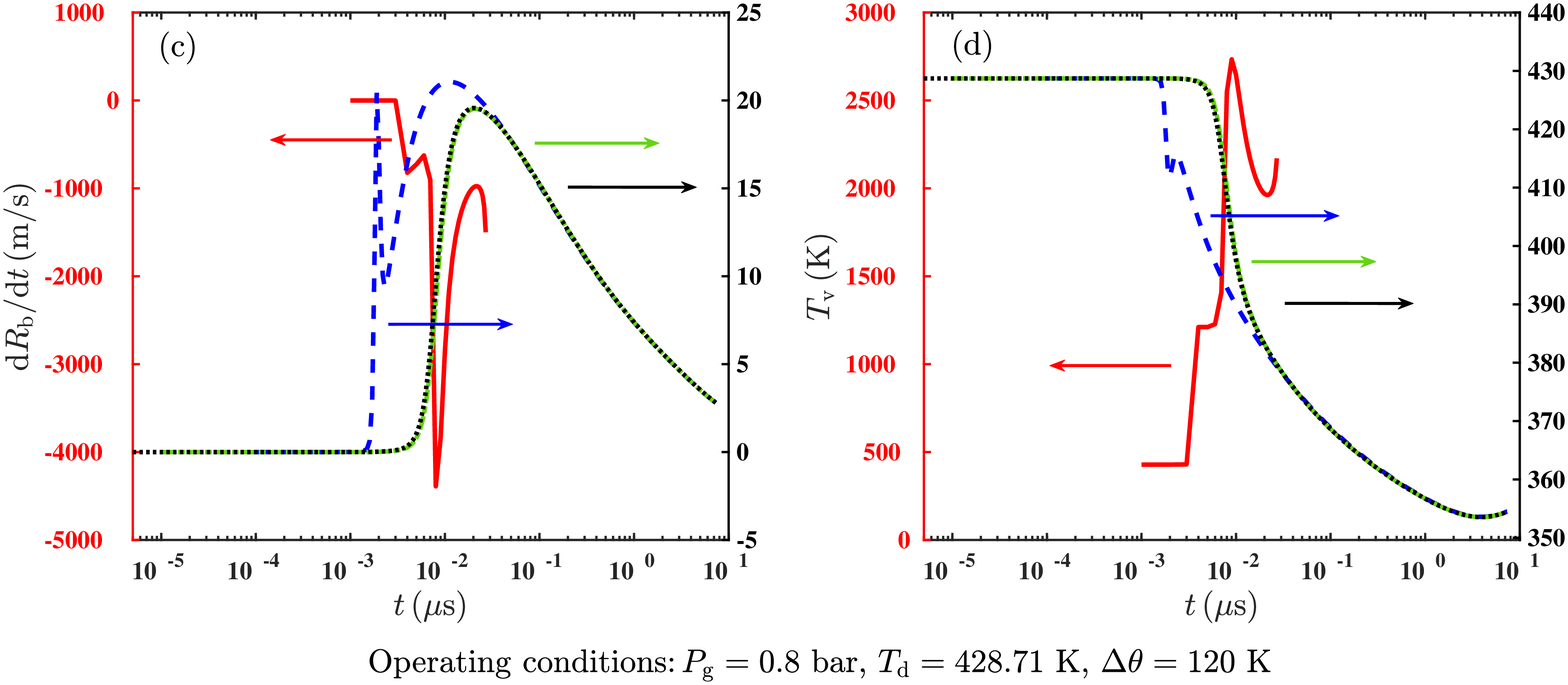} 
\vspace{-0.3 in}
\caption{Bubble growth rate (left column) and vapor temperature (right column) variation for $\mathrm{OME_1}$ microdroplets under various operating conditions and time-step sizes considering the bubble-bubble interactions.}
\label{Saha567wB}
\end{figure}
\reff{Saha567wB} shows the variation of bubble growth rate for different time-step sizes for $\OMEa$ microdroplets under two different operating conditions considering the bubble-bubble interactions. For $\Delta\theta=135$ K, it is observed that the time-step size of $\Delta t=1\times10^{-9}$ s causes the growth rate to become negative in the very beginning (see \reff{Saha567wB}a), which then causes an increasing vapor temperature, as shown in \reff{Saha567wB}b. The simulation is not continued further since the Newton-Raphson iteration estimates an unrealistically negative vapor temperature with the negative bubble growth rate in the next time-step. As the time-step size is decreased to $\Delta t=1\times10^{-10}$ s, the bubble evolves well in the very beginning. However, it shows a sudden jump in growth rate at around 7 $\times$$10^{-4}$ $\upmu$s, followed by a sharp decline at 9 $\times$$10^{-4}$ $\upmu$s. A sudden drop in vapor temperature can also be seen from \reff{Saha567wB}b due to a sudden rise in bubble growth rate. Nevertheless, the bubble growth rate increases further in time and subsequently results in decreasing $\Tv$. This fluctuating growth behavior does not appear to be physical~\citep{Saha2022}. The time-step sizes are then further reduced to $\Delta t=1\times10^{-11}$ s and $\Delta t=5\times10^{-12}$ s, which are found to eliminate the spurious fluctuations in the bubble growth rate curve, as obtained earlier with the larger time step sizes. All the physical phenomena associated with the bubble growth such as the surface-tension dominated extremely slow growth stage, followed by the rapid bubble surface acceleration and subsequent gradual deceleration are captured with the reduced $\Delta t$. Qualitatively, these results are in good agreement with previous publications~\citep{Robinson2002,Xi2017,Saha2022}. \reff{Saha567wB}c \& \reff{Saha567wB}d depict the bubble growth rate and vapor temperature variation for $\Delta\theta=120$ K. Similarly, $\Delta t=1\times10^{-9}$ s and 
$\Delta t=1\times10^{-10}$ s  result in unrealistic growth rates and vapor temperatures, whereas $\Delta t=1\times10^{-11}$ s and $\Delta t=5\times10^{-12}$ s correctly predict the evolution of the bubble. The numerical issue with the time-step size is also demonstrated using another highly volatile cryogenic e-fuel microdroplets of DME and the results are included in \ref{appE}.\par
\begin{figure}[!t]
\centering
\includegraphics[width=420pt]{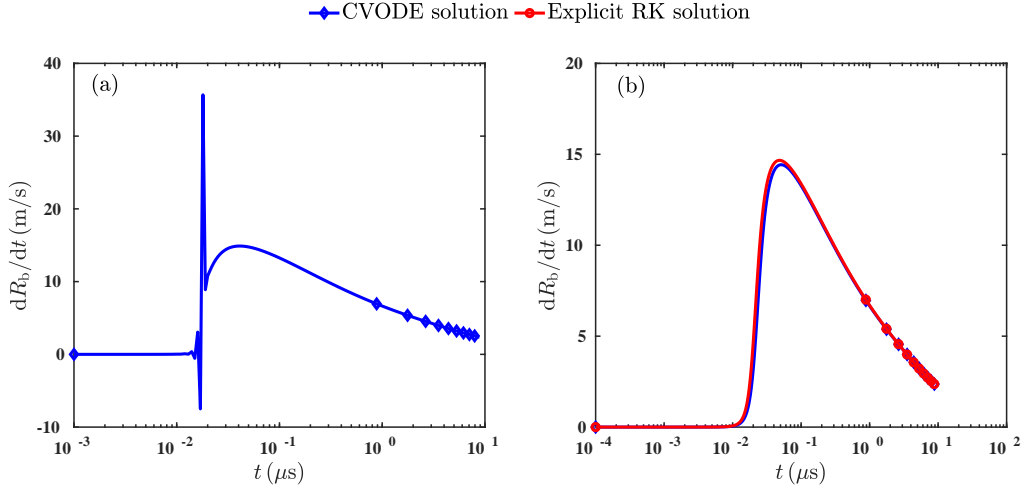}   
\caption{Bubble growth rate of $\OMEa$ microdroplets obtained using different solvers for (a) $\Delta t=1\times10^{-9}$ s and (b) $\Delta t=1\times10^{-10}$ s considering the bubble-bubble interactions at $\Pg=1.0$ bar, $\Td=404.86$ K, and $\Delta\theta=90$ K.}
\label{Saha19}
\end{figure}

 
The numerical results described above are obtained using the explicit RK solver as mentioned in \refs{CompStSolPrc}. An implicit ODE solver CVODE, which includes variable time-step, variable-order backward differentiation formulas, is also employed in this study to simulate the bubble growth characteristics. It is observed that the CVODE solver produces spurious fluctuations in the bubble growth rate curve near the beginning of the T growth stage. The simulation results obtained from the CVODE solver are shown here for $\Delta\theta=90$ K with $\Delta t=1\times10^{-9}$ s in \reff{Saha19}a. As the time-step size is reduced to $\Delta t=1\times10^{-10}$ s, the bubble growth rate is found to be well predicted by the CVODE solver similarly to the explicit RK solver, as shown in \reff{Saha19}b. However, due to its implicit nature, the computational cost associated with the CVODE solver is higher compared to the explicit RK solver. For example, the simulated case with the CVODE solver is run on a machine using the Intel Broadwell processor architecture on a single core for 1.2585 CPUh, whereas the same case with the explicit RK solver is run on the same machine on a single core for 0.0186 CPUh.
\subsection{Influence of bubble-bubble interactions on the bubble dynamics}
\begin{figure}[!t]
\centering
\includegraphics[width=420pt]{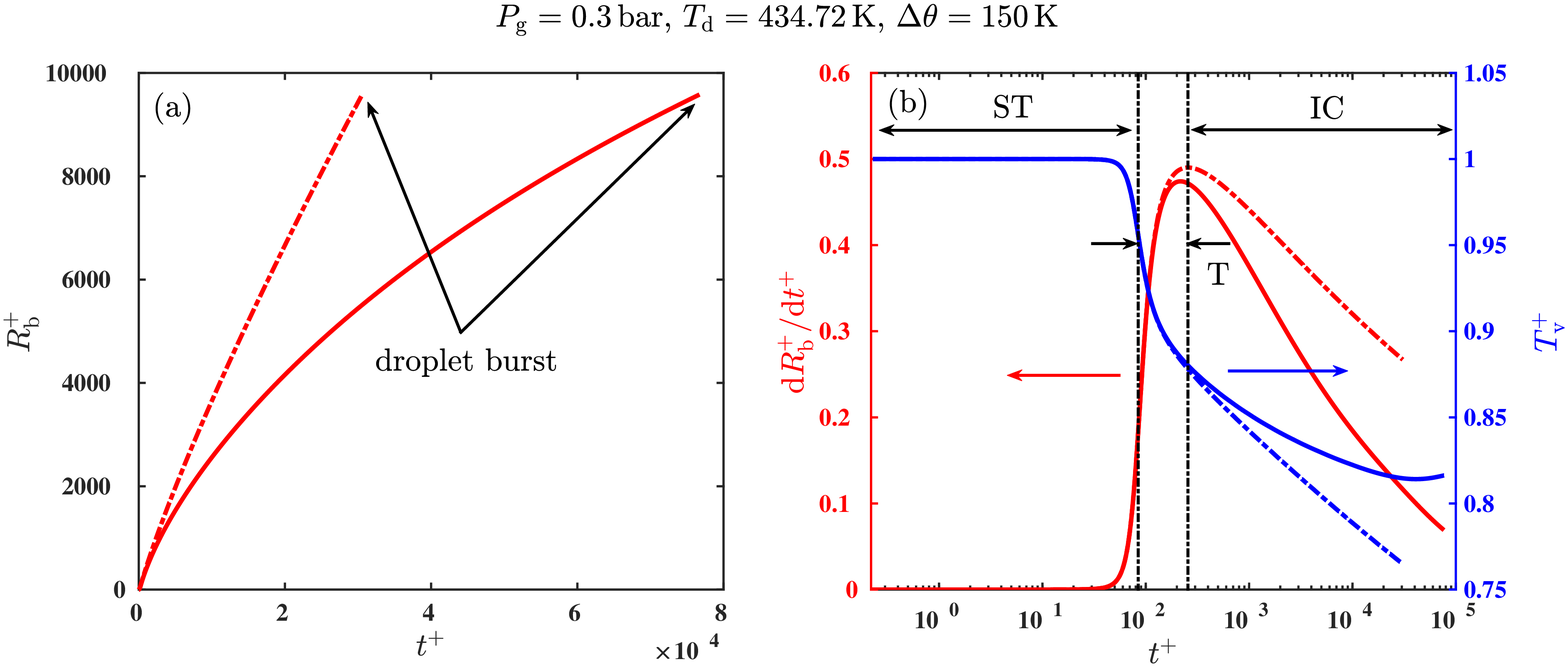}  
\vspace{-0.3 in}
\caption{Comparison of the bubble growth characteristics with (solid line) and without (dash-dotted line) bubble-bubble interactions for $\OMEa$ microdroplet at $\Pg=0.3$ bar, $\Td=434.72$ K, and $\Delta\theta=150$ K. Subfigure (a) shows the nondimensional bubble radius evolution over time. Subfigure (b) depicts the variation of the nondimensional bubble growth rate and vapor temperature. The initial vapor temperature is used as a reference temperature scale to nondimensionalize the vapor temperature $\Tv$.}
\label{Bubbub1}
\end{figure}

\begin{figure}[!h]
\centering
\includegraphics[width=420pt]{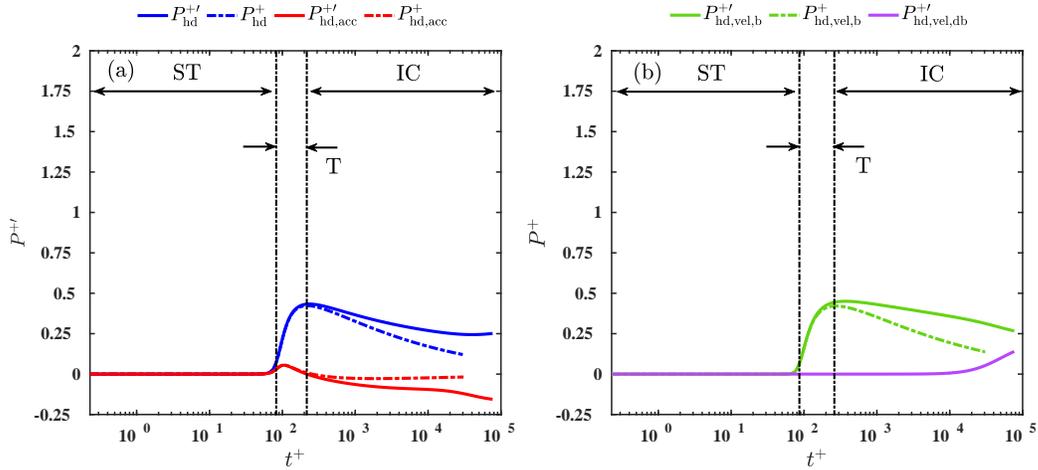}  
\caption{Comparison of the nondimensional pressure terms for the bubble growth with (solid line) and without (dash-dotted line) the bubble-bubble interactions for $\OMEa$ microdroplet at $\Pg=0.3$ bar, $\Td=434.72$ K, and $\Delta\theta=150$ K.}
\label{Bubbub2}
\end{figure}

The comparison between the bubble growth characteristics of $\OMEa$ microdroplets with and without bubble-bubble interactions is shown in \reff{Bubbub1} for $\Pg=0.3$ bar, $\Td=434.72$ K, and $\Delta\theta=150$ K. It can be seen that the bubble growth characteristics are significantly influenced by the bubble-bubble interactions once the bubbles surpass the T stage and enter into the IC growth stage. \reff{Bubbub1}a depicts that the bubble-bubble interactions cause much slower expansion of the bubbles. This is because the additional pressure force induced by the surrounding growing bubbles on the target bubble surface markedly reduces its growth rate in the IC growth stage, as illustrated in \reff{Bubbub1}b. It is also observed from \reff{Bubbub1}b that the rate at which the vapor temperature decreases considering the bubble-bubble interactions decreases substantially in this growth stage and subsequently reduces the positive thermal feedback at the bubble surface, which is proportional to $\DeltaTplus=\Tdplus-\Tvplus$. The higher hydrodynamic pressure emerging from the bubble-bubble interactions in the IC growth stage is eventually the reason behind this phenomenon, which is depicted in \reff{Bubbub2}. \par
From \reff{Bubbub1}b, it is interesting to note that the vapor temperature curve for the bubble growth with bubble-bubble interactions shows a gradual rise near the onset of droplet bursting. This phenomenon can be explained as follows: the mass within the bubbles increases in time as the surrounding superheated liquid becomes vapor due to the continuous energy flux at the bubble surface. The increasing mass causes a build-up of vapor pressure inside the bubbles. The bubbles then grow in size such that this excess of pressure can be relaxed. Since the vapor inside the bubbles is assumed to be saturated, a reduction in $\Pv$ also lowers the vapor temperature $\Tv$. However, for the bubble growth considering the bubble-bubble interactions, a gradual rise in $\Phdplusprime$ near the onset of the droplet bursting suppresses the bubble growth, and subsequently the bubble surface acceleration (as shown in \reff{Bubbub2}a) to such an extent that the vapor pressure continues to build up due to the slower growth of the vapor bubbles, and thus leading to an increase in vapor temperature near the onset of droplet bursting, as shown in \reff{Bubbub1}b. The faster droplet expansion in this regime eventually increases $\Phdveldbplusprime$ (as can be seen from \reff{Bubbub2}b) in the later period of IC growth, which in turn, causes the total hydrodynamic pressure, $\Phdplusprime$ to increase (note that the slope of $\Phdprime$ curve changes from negative to positive in the IC growth stage). Whereas for the bubble growth without the bubble-bubble interactions, $\Phdplus$ continues to decrease in the IC growth stage due to the absence of the additional pressure force term, $\Phdveldbplusprime$. Thus, the vapor temperature also shows a continuous decline in time, as shown by the blue dash-dotted line in \reff{Bubbub1}b. Due to the slower growth of the vapor bubbles, the onset of droplet bursting is also considerably delayed for the bubble growth with bubble-bubble interactions. The variation of the nondimensional pressure terms due to the viscous and surface tension forces is not shown as the bubble-bubble interactions are found to have negligible influence on these forces.\par
\begin{figure}[!h]
\centering
\includegraphics[width=420pt]{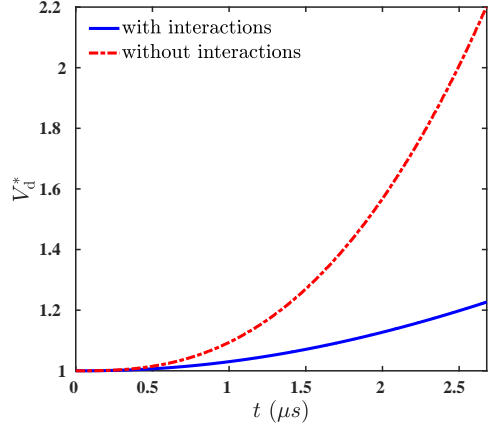}  
\vspace{-0.3 in}
\caption{Variation of the dimensionless volume of the $\OMEa$ fuel droplet for the bubble growth with (solid line) and without (dash-dotted line) the bubble-bubble interactions at $\Pg=0.3$ bar, $\Td=434.72$ K, and $\Delta\theta=150$ K. The initial volume of the liquid droplet is used as a reference scale to nondimensionalize the instantaneous droplet volume, $V_\text{droplet}$.}
\label{Bubbub3}
\end{figure}

 
The dimensionless volume of the $\OMEa$ fuel droplet is shown in \reff{Bubbub3} for the bubble growth with and without bubble-bubble interactions up to 2.67 $\upmu$s, which marks the onset of droplet bursting for the later case. It is observed that the volumetric expansion of the liquid droplet is significantly smaller in presence of bubble-bubble interactions. It can be seen that the droplet volume has increased by only 23\% after 2.67 $\upmu$s when considering bubble-bubble interactions. Conversely, the case without bubble-bubble interactions leads to an increase in droplet volume by approximately 120\% within the same time interval. \par
\subsection{Scaling law for superheated microdroplets}
The scaling relationships of \cite{Lien1969} (\refe{eqn:AB}) work well only for the later bubble growth stages when the bubble surpasses the T stage and enters into the IC or TD growth stage~\citep{Lee1996}. This is expected because the effect of surface tension, which remains a dominating factor during the first two growth stages~\citep{Robinson2002,Saha2022}, is not considered in the \cite{Lien1969} formulation. The \cite{Prosperetti1978} scaling law (\refe{Pros}) also shows a behavior similar to the \cite{Lien1969} scaling law because it assumes a linear variation of saturated vapor pressure with temperature and constant physical properties in the formulations to describe the scaled bubble growth behavior.\par
\begin{figure}[h]
\centering
\includegraphics[width=420pt]{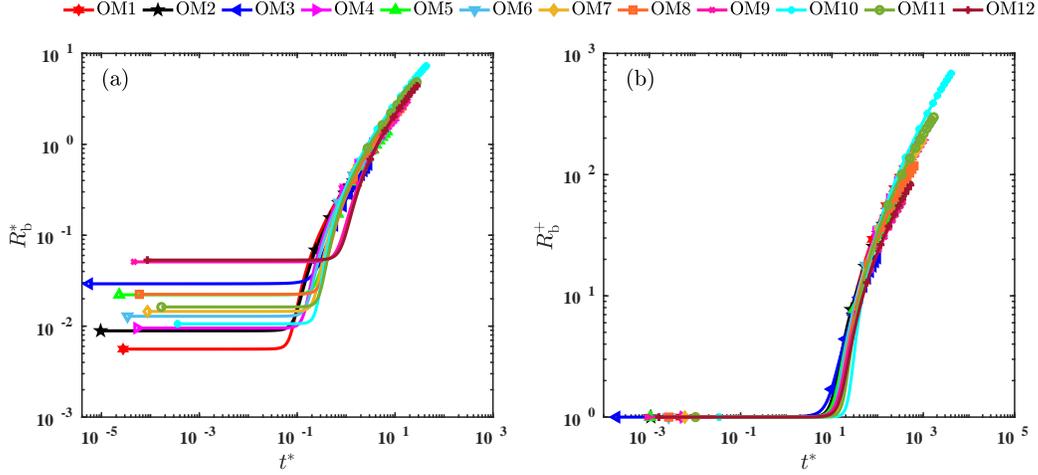}     
\caption{Nondimensional radius of the vapor bubbles in superheated $\OMEa$ microdroplets scaled with \refe{eqn:AB} (a) and \refe{eqn:scl_qnt1} (b) scaling law under subatmospheric pressure conditions, as listed in \reft{OMcases}. The bubble radii are plotted in solid lines with various types of markers associated with different test cases considered for analyzing the behavior of the scaling laws.}
\label{Saha9}
\end{figure}

In real engine conditions, liquid jets or droplets will not remain intact for a long period. They will rupture within a very short period (in the order of microseconds) due to the rapid growth of multiple vapor bubbles. Thus, the scaling laws, \refe{eqn:AB} and \refe{Pros}, which work well in later bubble growth stages when the bubbles are grown sufficiently large beyond their critical radius, will not be applicable for the realistic engine conditions. This is illustrated in \reff{Saha9}a, where the temporal variations of nondimensional bubble radii of $\OMEa$ microdroplets are plotted using the \cite{Lien1969} scaling law considering the bubble-bubble interactions for various operating conditions, as listed in \reft{OMcases}. Hence, the need arises for a more general scaling law, which reasonably models not only the later IC and TD growth stages but also the initial ST growth stage. \par
\begin{table}[t]
  \begin{center}
  \def~{\hphantom{0}}
  \begin{tabular}{l@{\quad}l@{\quad}c@{\quad}l@{\quad}l@{\quad}r@{\quad}r}
  \toprule
      \textbf{Case} & \hskip0.5cm\text{$\Pl \left(\mathrm{bar}\right)$} & \hskip0.5cm\textbf{$\Td \left(\mathrm K\right)$}          & \hskip0.5cm\textbf{$\Delta\theta \left(\mathrm K\right)$} & \hskip0.5cm\textbf{$\Rc \left(\mathrm{\upmu m}\right)$}     & \hskip0.5cm\textbf{$A \left(\mathrm{m/s}\right)$}     & \hskip0.5cm\textbf{$B \left(\mathrm{m/\sqrt s}\right)$}   \\ [3pt]
  \midrule
OM1         & \hskip0.5cm 0.3                                                                 & \hskip0.5cm 331.72       & \hskip0.5cm 47  & \hskip0.5cm 0.22       & \hskip0.5cm 10.95 & \hskip0.5cm 0.021                  \\ [3pt]
OM2           & \hskip0.5cm 0.3                                                        & \hskip0.5cm 311.72       & \hskip0.5cm 27  & \hskip0.5cm 0.64        & \hskip0.5cm 6.88 & \hskip0.5cm 0.022               \\ [3pt]
OM3           & \hskip0.5cm 0.3                                                        & \hskip0.5cm 296.72      & \hskip0.5cm 12  & \hskip0.5cm 2.08      & \hskip0.5cm 3.97 & \hskip0.5cm 0.017                \\ [3pt]
OM4           & \hskip0.5cm 0.5                                                      & \hskip0.5cm 336.63       & \hskip0.5cm  40 & \hskip0.5cm 0.20  & \hskip0.5cm 11.32       & \hskip0.5cm 0.015                \\ [3pt]
OM5           & \hskip0.5cm 0.5                                                       & \hskip0.5cm 316.63      & \hskip0.5cm 20                                                                        & \hskip0.5cm 0.65  & \hskip0.5cm 6.73 & \hskip0.5cm 0.014    \\ [3pt]
OM6         & \hskip0.5cm 0.5                                                                 & \hskip0.5cm 326.63       & \hskip0.5cm 30  & \hskip0.5cm  0.34       & \hskip0.5cm 9.0 & \hskip0.5cm 0.015                  \\ [3pt]
OM7           & \hskip0.5cm 0.7                                                        & \hskip0.5cm 340.16       & \hskip0.5cm 35  & \hskip0.5cm 0.19        & \hskip0.5cm 11.45 & \hskip0.5cm 0.012               \\ [3pt]
OM8           & \hskip0.5cm 0.7                                                       & \hskip0.5cm 330.16      & \hskip0.5cm 25  & \hskip0.5cm 0.34       & \hskip0.5cm 8.93 & \hskip0.5cm 0.012                \\ [3pt]
OM9           & \hskip0.5cm 0.7                                                       & \hskip0.5cm 320.16       & \hskip0.5cm 15 & \hskip0.5cm 0.70  & \hskip0.5cm 6.37       & \hskip0.5cm 0.009                \\ [3pt]
OM10           & \hskip0.5cm 0.9                                                        & \hskip0.5cm 368.92      & \hskip0.5cm 57                                                                        & \hskip0.5cm 0.06  & \hskip0.5cm 18.84 & \hskip0.5cm 0.010    \\[3pt]
OM11           & \hskip0.5cm 0.9                                                        & \hskip0.5cm 349.92       & \hskip0.5cm 38 & \hskip0.5cm 0.13  & \hskip0.5cm 13.36       & \hskip0.5cm 0.010               \\ [3pt]
OM12           & \hskip0.5cm 0.9                                                       & \hskip0.5cm 328.92      & \hskip0.5cm 17                                                                       & \hskip0.5cm 0.47  & \hskip0.5cm 7.60 & \hskip0.5cm 0.008    \\
  \bottomrule
  \end{tabular}
  \caption{Simulation test cases for $\OMEa$ microdroplets.}
  \label{OMcases}
  \end{center}
\end{table}
\reff{Saha9}b shows the variation of dimensionless bubble radii plotted with \refe{eqn:scl_qnt1} considering the bubble-bubble interactions for the similar pressure and temperature ranges, as shown in \reft{OMcases}. It can be seen from \reff{Saha9}b that all the dimensionless radius curves merge into a single curve of value unity in the regime of ST growth. In the later growth stages, however, the scaled radius-time curves do not perfectly collapse into a single curve. Thus, it is difficult to describe a reasonably accurate universal bubble growth behavior with a relation of the form
\begin{equation}
\Rplus = f\left(\tplus\right).
\end{equation} 
\subsection{Dimensional analysis of the modified RPE}
In this section, first, a dimensional analysis of the modified RPE (\refe{RPEndwB}) is performed to determine the magnitude of the different nondimensional force terms acting on the bubble surface for Reynolds numbers of different orders of magnitude. Based on the dimensional analysis, a semi-analytical solution for the bubble growth is then derived in terms of nondimensional parameters considering the bubble-bubble interactions to mitigate the numerical issues described in \refs{Limit}. Finally, the proposed semi-analytical solution is validated with the numerical solution of the modified RPE.\par
\begin{figure}[t]
\centering
\includegraphics[width=405pt]{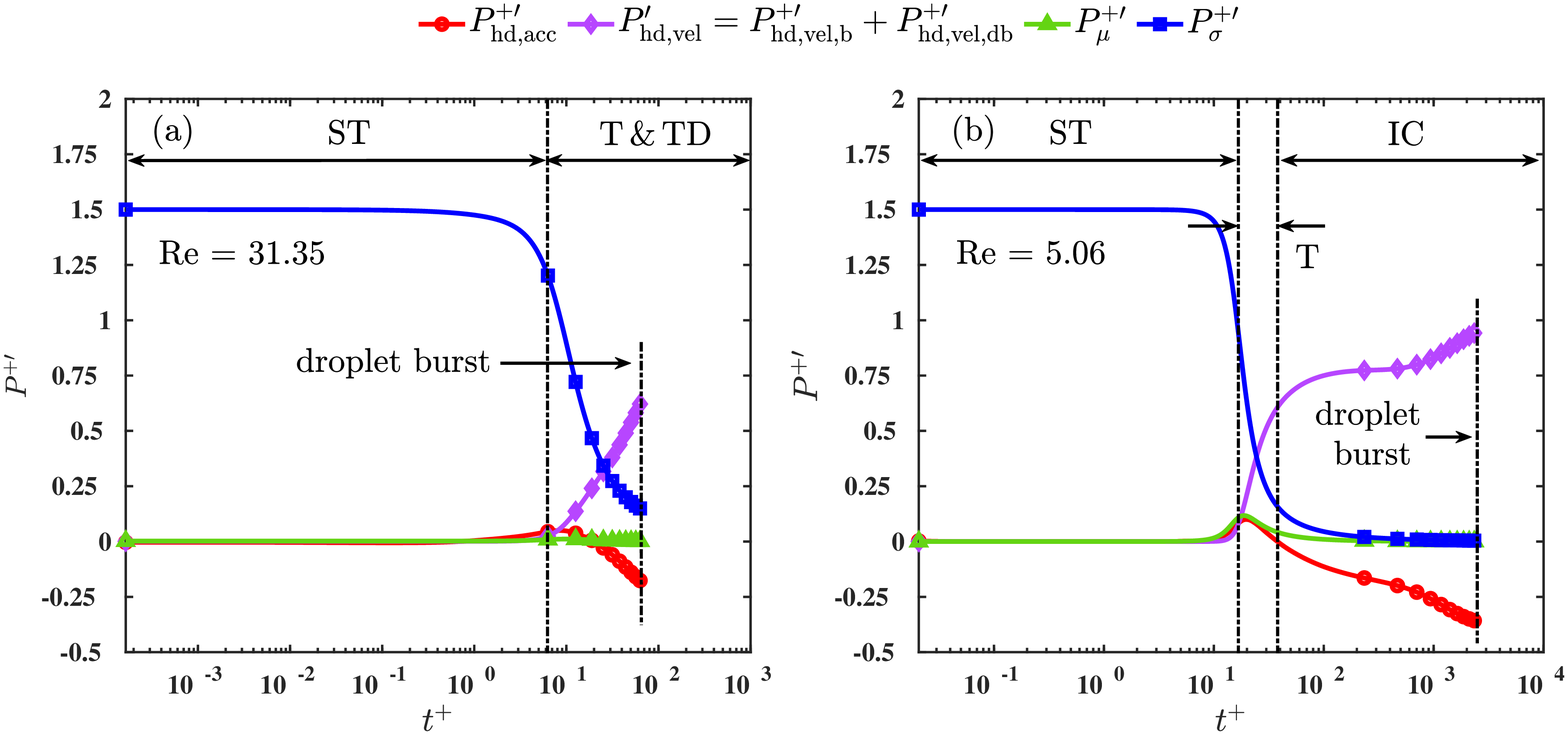}   
\includegraphics[width=405pt]{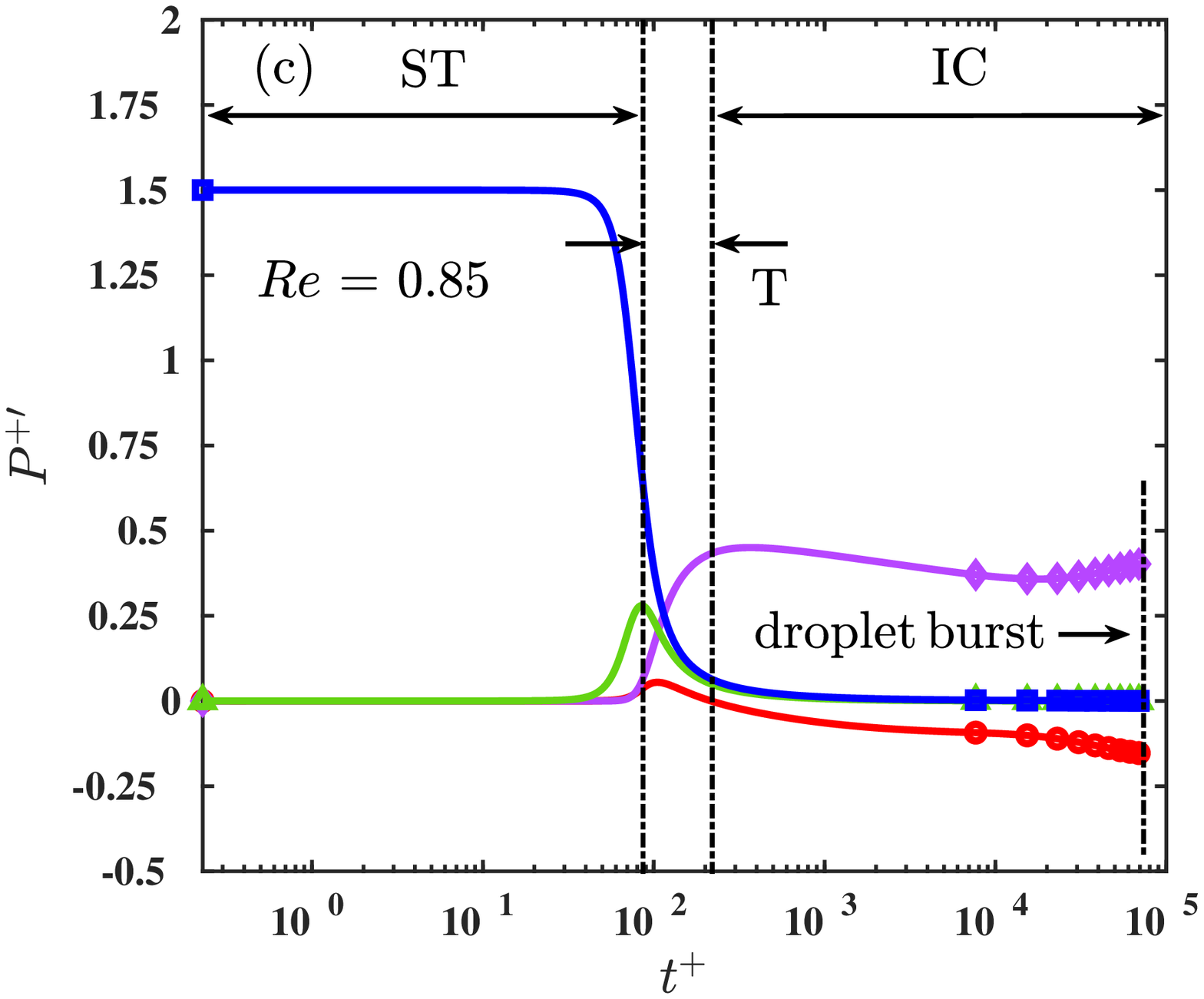}   
\vspace{-0.6 in}
\caption{Comparison of the different nondimensional pressure terms of \refe{RPEndwB} as a function of nondimensional time for three different Reynolds numbers of (a) 31.35, (2) 5.06, and (c) 0.85, considering bubble-bubble interactions in $\OMEa$ microdroplets for operating conditions listed in \reft{Recases}. The pressure terms are plotted in solid lines with different types of markers in order to distinguish the contributions from the inertia, viscous, and surface tension forces}.
\vspace{-0.18 in}
\label{Saha10wB}
\end{figure}

\reff{Saha10wB} illustrates the variation of $\Pplusprime$ for three different Re with bubble-bubble interactions. The physical parameters and their values corresponding to the different Reynolds number cases are listed in \reft{Recases}. It is to be noted that the definition of the Re shown in \refe{ReWe} allows a maximum value of Re in the order of the magnitude of 2, which corresponds to a very low superheating degree case ($\Delta\theta\approx\mathcal{O}(10^{-2}\,\text{K})$). In this study, we have considered three different Re of 31.35, 5.06, and 0.85, termed as high Reynolds number (`HRe'), moderate Reynolds number (`MRe'), and low Reynolds number (`LRe'), respectively. 
\begin{table}[h]
  \begin{center}
\def~{\hphantom{0}}
  \begin{tabular}{l@{\quad}c@{\quad}c@{\quad}c@{\quad}r@{\quad}r}
  \toprule
      \textbf{Parameter}                      &  \begin{tabular}[c]{@{}c@{}} \textbf{Case \#1} \\\textbf{High Re $\left(\mathrm {HRe}\right)$}\end{tabular} &   \begin{tabular}[c]{@{}c@{}} \textbf{Case \#2} \\ \textbf{Moderate Re $\left(\mathrm {MRe}\right)$}\end{tabular} & \begin{tabular}[c]{@{}c@{}}  \textbf{Case \#3} \\ \textbf{Low Re $\left(\mathrm {LRe}\right)$}\end{tabular} \\
  \midrule
$\Pl \left(\mathrm{bar}\right)$                                 & 0.3 & 0.3 & 0.3          \\
$\Td \left(\mathrm{K}\right)$                              & 290.72 & 355.72 & 434.72         \\
$\Delta\theta \left(\mathrm{K}\right)$                 & 6 & 71  & 150       \\ 
$\Rc \left(\mathrm{\upmu m}\right)$   & 4.83 &    0.0802    & 0.004 \\ 
$A \left(\mathrm{m/s}\right)$                  & 2.64  & 16.62 &  44.49    \\
$B \left(\mathrm{m/s^{\frac{1}{2}}}\right)$     & 0.011 & 0.017  & 0.008        \\
Re & 31.35 & 5.06  & 0.85\\
  \bottomrule
  \end{tabular}
  \caption{Simulation test cases of $\OMEa$ microdroplets for validating the simplified semi-analytical solutions.}
  \label{Recases}
  \end{center}
\end{table}
It can be seen from \reff{Saha10wB}a that for the case of `HRe', the order of magnitude of $\Pmuplusprime$ is negligible throughout the whole bubble lifetime compared to the other terms. However, with decreasing Re, the influence of $\Pmuplusprime$ becomes important already in the later period of the ST growth stage, as shown in \reff{Saha10wB}c for `LRe'. The order of magnitude of $\Phdaccbplusprime$ starts to become important towards the end of the ST growth stage and becomes negative in the transition stage with gaining importance for later times. Nevertheless, as the Re is reduced, the rate of decrease in $\Phdaccbplusprime$ also decreases. Thus, it can be concluded that $\Phdaccbplusprime$ becomes less important in determining the bubble growth characteristics compared to $\Phdvelplusprime$ as the Re is reduced from 31.35 to 0.85. Since $\Phdaccbplusprime$ has a tendency to become less significant with decreasing Re, we neglect this term in the following to derive an approximate semi-analytical solution of \refe{RPEndwB}. The effect of neglecting $\Phdaccbplusprime$ on bubble growth behavior will be discussed later in detail. Neglecting $\Phdaccbplusprime$, \refe{RPEndwB} becomes
\begin{equation}\label{RPEnd34}
   \begin{split}
     \Pvplus-\Pgplus=\rholplus\left(\frac{3}{2}+4 \pi \Rdplustwo \Ndenplus \Rbplus\right)\left(\frac{\mathrm{d} \Rbplus}{\mathrm{~d} \tplus}\right)^{2}+\frac{4}{\Rbplus} \frac{\mulplus}{\text{Re}}\frac{\mathrm{d} \Rbplus}{\mathrm{~d} \tplus }+4 \pi \Ndenplus \rholplus \Rdplus \Rbplustwo \left(\frac{\mathrm{d} \Rdplus}{\mathrm{~d} \tplus}\right) \left(\frac{\mathrm{d} \Rbplus}{\mathrm{~d} \tplus}\right)+\frac{2\sigmaplus}{\text{We}}\frac{1}{\Rbplus}.
  \end{split}
\end{equation}
Defining
\begin{equation}\label{A}
    X = \rholplus\left(\frac{3}{2}+4 \pi \Rdplustwo \Ndenplus \Rbplus\right),
\end{equation}
\begin{equation}\label{B}
    Y = 4\left\{\frac{\mulplus}{\Rbplus \text{Re}}+ \pi \Ndenplus \rholplus \Rdplus \Rbplustwo \left(\frac{\mathrm{d} \Rdplus}{\mathrm{~d} \tplus}\right)\right\},
\end{equation}
\begin{equation}\label{C}
    Z = \left\{\frac{2\sigmaplus}{\text{We}}\frac{1}{\Rbplus}-\left(\Pvplus-\Pgplus\right)\right\}, 
\end{equation}
\refe{RPEnd34} becomes
\begin{equation}\label{RPEnd6}
   \begin{split}
     X\left(\frac{\mathrm{d} \Rbplus}{\mathrm{~d} \tplus}\right)^{2}+Y\left(\frac{\mathrm{d} \Rbplus}{\mathrm{~d} \tplus}\right)+Z=0.
  \end{split}
\end{equation}
Treating \refe{RPEnd6} as a quadratic equation for $\mathrm d \Rbplus/\mathrm d \tplus$, it can be analytically solved for $\mathrm d \Rbplus/\mathrm d \tplus$ as
\begin{equation}\label{RPEnd5}
   \begin{split}
     \left(\frac{\mathrm{d} \Rbplus}{\mathrm{~d} \tplus}\right)=\frac{-Y\pm\sqrt{Y^2-4XZ}}{2X}.
  \end{split}
\end{equation}
Replacing $X$, $Y$, and $Z$ from \refe{A}, \refe{B}, and \refe{C}, respectively, into \refe{RPEnd5} and assuming the temporal variation of $\rholplus$, $\mulplus$, and $\sigmaplus$, to be negligible, one obtains
\begin{equation}\label{RPEnd7}
    \begin{split}
        \left(\frac{\mathrm{d} \Rbplus}{\mathrm{~d} \tplus}\right)=-\frac{2\left\{\frac{1}{\Rbplus \text{Re}} +\pi \Ndenplus\Rdplus \Rbplustwo \left(\frac{\mathrm{d} \Rdplus}{\mathrm{~d} \tplus}\right)\right\}}{\frac{3}{2}+4 \pi \Rdplustwo \Ndenplus \Rbplus}+\left[4\left\{\frac{\frac{1}{\Rbplus \text{Re}}+\pi \Ndenplus\Rdplus \Rbplustwo \left(\frac{\mathrm{d} \Rdplus}{\mathrm{~d} \tplus}\right) }{\frac{3}{2}+4 \pi \Rdplustwo \Ndenplus \Rbplus}\right\}^2 - \frac{\left\{\frac{2}{\Rbplus \text{We}}-\left(\Pvplus-\Pgplus\right)\right\} }{\left(\frac{3}{2}+4 \pi \Rdplustwo \Ndenplus \Rbplus\right)}\right]^{1/2}.
    \end{split}
\end{equation}
The negative root does not have physical significance in the context of vapor bubble growth in superheated microdroplets, and therefore is neglected in this study. \refe{RPEnd7} describes the simplified semi-analytical solution for the bubble growth, which needs to be numerically integrated to obtain the bubble radius. The assumptions considered to derive \refe{RPEnd7} are summarized as: (1) spherical symmetric bubble; (2) $\Rb/\Rd\ll 1$; (3) $\Ndenplus=$ constant; (4)~$\Phdaccbplusprime\approx 0$, and (5) $\rholplus\approx$ $\mulplus\approx$ $\sigmaplus\approx$ 1.\par
\begin{figure}[h]
\centering
\includegraphics[width=400pt]{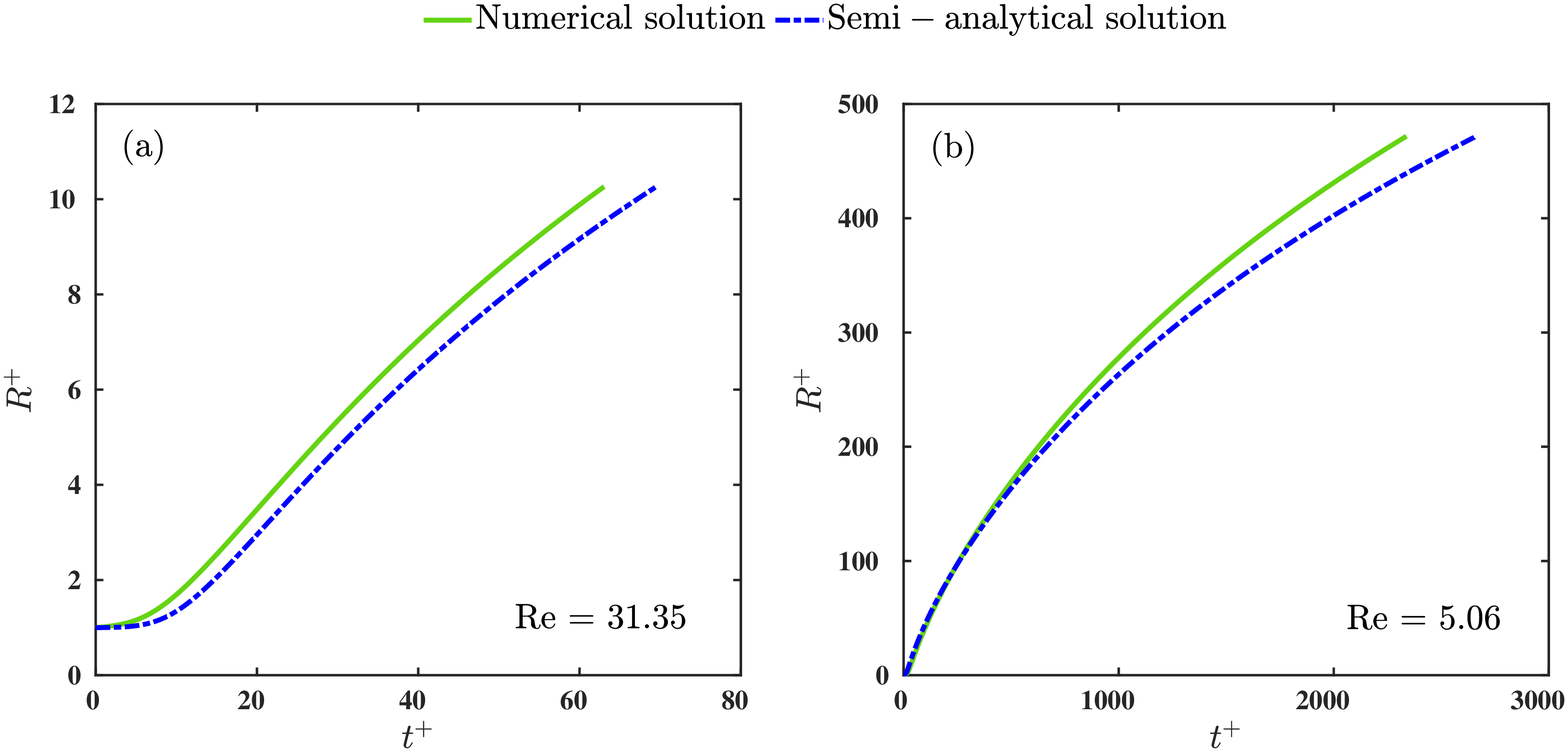}    
\includegraphics[width=400pt]{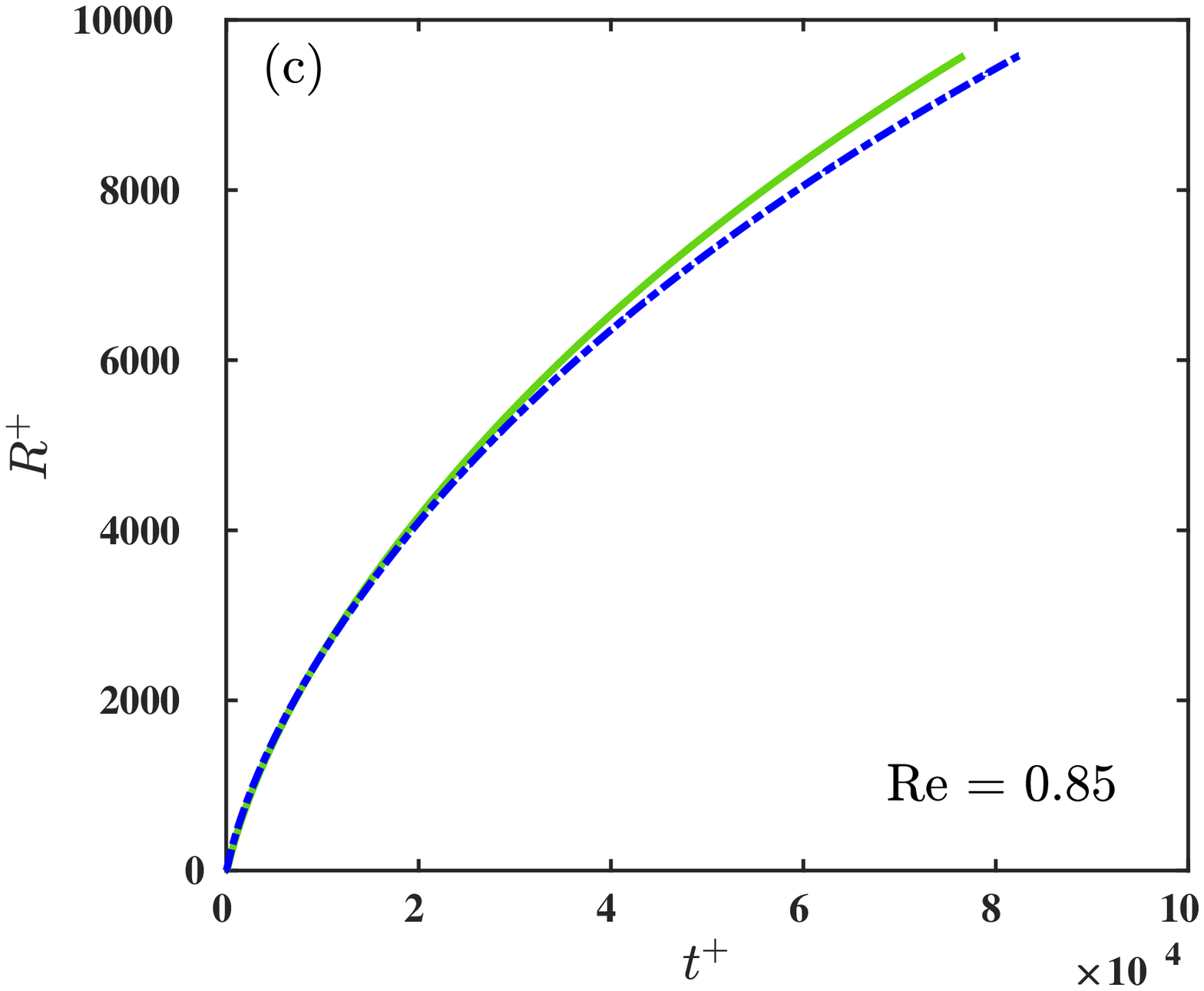}    
\vspace{-0.5 in}
\caption{Comparison of the dimensionless radius of the superheated $\OMEa$ vapor bubbles obtained from the numerical solution of \refe{RPEndwB}) with the simplified semi-analytical solution of \refe{RPEnd7} for (a) Re = 31.35 (Case `HRe'), (b) Re = 5.06 (Case `MRe'), and (d) Re = 0.85 (Case `LRe'). The detailed physical parameters of all the three simulated test cases are listed in \reft{Recases}.}
\label{Saha1314wB}
\end{figure}

\reff{Saha1314wB} compares the numerically obtained nondimensional radius-time behavior of the $\OMEa$ vapor bubble with the dimensionless semi-analytical solution for three Reynolds numbers of 31.35 (\reff{Saha1314wB}a), 5.06 (\reff{Saha1314wB}b), and 0.85 (\reff{Saha1314wB}c), as listed in \reft{Recases}. The three different Re cases are chosen based on the different superheating degrees ranging from very low ($\theta$ = 6 K) to super-high ($\theta$ = 150 K) values. We have considered the numerical solution of the modified RPE (\refe{RPEndwB}) as the reference to validate the proposed semi-analytical solution. It is observed from \reff{Saha1314wB} that the semi-analytical solution underpredicts the bubble growth behavior for all the investigated cases. This is expected due to the neglection of the acceleration term in the derivation of \refe{RPEnd7}. Nevertheless, it can be seen that the underprediction reduces with decreasing Re. This is because the acceleration term becomes less significant with decreasing Re as also revealed by the analysis of the order of magnitude of the different pressure terms in \reff{Saha10wB}. Overall, it can be concluded that the simplified nondimensional semi-analytical solution provides a good approximation of the bubble evolution for the whole bubble lifetime.\par
It has to be noted that the derived semi-analytical solution does not depend on the source (e.g., turbulence, throttling effect, etc.) and location of inception (e.g., inside or outside of the injector nozzle) of the vapor bubble. As soon as a bubble nucleates in a superheated liquid medium, it can be used to reasonably predict the bubble evolution until the liquid jet (or droplet) bursts even with the larger time-step sizes compared to the numerical approach.
\subsection{Computational cost}
\begin{figure}[t]
\centering
\includegraphics[width=420pt]{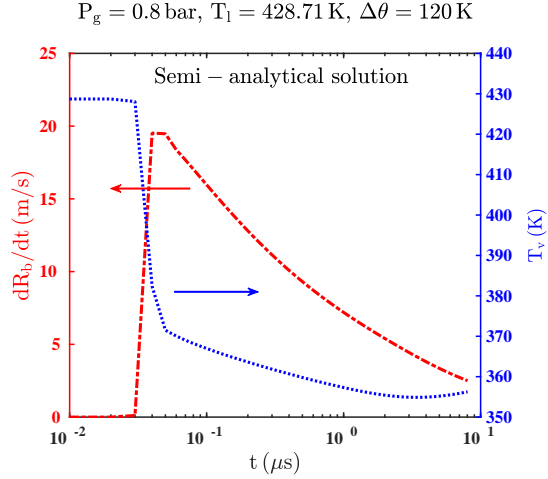}    
\caption{Bubble growth characteristics of $\OMEa$ microdroplets considering the bubble-bubble interactions predicted by the proposed semi-analytical solution for bubble growth (\refe{RPEnd7}) with $\Delta t=1\times10^{-8}$s.}
\vspace{-0.1 in}
\label{Saha11}
\end{figure}

The proposed semi-analytical solution is found to accurately predict the bubble growth behavior with a time-step size of $\Delta t=1\times10^{-8}$ s, whereas a much smaller time-step size is required by the numerical solver (as described in \refs{Limit}), thus substantially increasing the required number of time-steps. In order to demonstrate the capability of the derived semi-analytical solution, the numerical and semi-analytical predictions of the bubble growth characteristics for $\OMEa$ microdroplet are shown in \reff{Saha11} for $\Pg=0.8$ bar, $\Td=428.71$ K, and $\Delta\theta=120$ K with $\Delta t=1\times10^{-8}$ s. It is observed in \reff{Saha11} that the semi-analytical solution accurately captures the growth characteristics of the vapor bubbles regardless of the large time-step size, whereas the numerical approach results in an unphysical bubble growth rate and vapor temperature, as described in \refs{Limit}.\par
Five different cases with superheating degrees ranging from 80 K (Case `A-80') to 150 K (Case `E-150') are investigated to illustrate the reduction of the computational cost with the derived semi-analytical formulation. In this work, all the simulations are performed on machines using the Intel Broadwell processor architecture. The simulation parameters are listed in \reft{Cost}. The semi-analytical solution is found to accelerate the computing speed significantly for all the investigated cases, as shown by factor x in \reft{Cost}.  Thus, for large-scale spray simulations with millions of microdroplets~\citep{Davidovic2017}, the semi-analytical solution would be very beneficial to speed up the whole computation process.
\begin{table}[!b]
  \begin{center}
\def~{\hphantom{0}}
  \begin{tabular}{l@{\quad}c@{\quad}c@{\quad}c@{\quad}c@{\quad}c@{\quad}c@{\quad}c@{\quad}c@{\quad}r}
  \toprule
      \textbf{Case}                      &  $\Pl \left(\mathrm{bar}\right)$  &   $\Td \left(\mathrm{K}\right)$    & $\Delta\theta \left(\mathrm{K}\right)$ & $\Delta t_\text{num} \left(\text s\right)$ & $\Delta t_\text{anlt} \left(\text s\right)$ & $\Tnm \left(\text{CPUh}\right)$ & $\Tan \left(\text{CPUh}\right)$ & \textbf{x}\\ [3pt]
  \midrule
A-80                                & 0.3 & 364.72 & 80 & $1\times10^{-9}$   & $1\times10^{-8}$   & 0.0021 & $1.2792\times10^{-4}$ & 16.42   \\
B-100                    & 0.3 & 384.72 & 100 & $1\times10^{-9}$  & $1\times10^{-8}$ & 0.0016 & $9.8798\times10^{-5}$ & 16.19\\ 
C-120      & 0.8 &   428.71   & 120  & $1\times10^{-11}$ & $1\times10^{-8}$ & 0.6581 & $3.7546\times10^{-4}$ & 1752.78\\ 
D-135        & 0.6 & 436.20 & 135 & $1\times10^{-11}$   & $1\times10^{-8}$  & 0.6664 &  $3.4923\times10^{-4}$ & 1908.20 \\
E-150        & 0.3 & 434.72 & 150 & $1\times10^{-11}$   & $1\times10^{-8}$  &  0.1444 &  $3.4146\times10^{-4}$ & 422.89 \\
  \bottomrule
  \end{tabular}
  \caption{Simulation test cases of $\OMEa$ microdroplets for demonstrating the computational cost reduction with the proposed semi-analytical solution. Subscripts `num' and `anlt' denote numerical and analytical solver, respectively.}
  \label{Cost}
  \end{center}
\end{table}
\subsection{Limitations of the proposed ROM}
Several assumptions were made in the derivation of the modified RPE (\refe{RPE5}), such as constant bubble number density, monodisperse and spherically symmetric vapor bubbles, and sufficiently large inter-bubble spacing, which may not always hold true in comparison to the practical scenario. It has to be noted that while the assumptions are being made, the proposed model is closer to reality than completely neglecting the interactions. A more advanced subgrid-scale model for bubble growth may be derived by relaxing these assumptions. For example, polydisperse vapor bubbles arranged randomly in a superheated microdroplet could be considered. However, such an advanced modeling approach requires solving the RPE for each of the vapor bubbles present in the superheated droplet. This is also likely to increase the computational cost for large-scale sprays due to the presence of millions of microdroplets with multiple polydisperse vapor bubbles. Although it is necessary to study the dynamics of the droplet-bubble system with the above-mentioned advanced modeling strategy, the actual need for such models in predicting global spray characteristics should be determined based on the comparative analysis of cost-accuracy trade-off for different subgrid-scale bubble growth models with varying degrees of simplification.

\section{Conclusions} \addvspace{10pt}
\label{Conc}
A new ROM for vapor bubble growth considering the bubble-bubble interactions in superheated microdroplets is proposed in this study. The model performance is compared with the DNS performed by \cite{Dietzel2019a}. The present reduced-order bubble-growth model was shown to capture the trend in considerably smaller volumetric droplet expansion relative to the single isolated bubble case. It was found that the bubble-bubble interactions significantly influence the bubble growth characteristics once the bubble surpasses the transition growth stage. The additional inertial force, $\Pinertia$, emerging from the bubble-bubble interactions was found to markedly reduce the bubble growth rate, and thus, significantly delay the onset of droplet bursting.\par
The numerical challenges associated with the proposed bubble growth model for the accurate prediction of the vapor bubble growth process were highlighted for highly volatile liquid fuels using $\mathrm{OME_x}$ as a generic example. It was found that for certain operating conditions, the numerical approach requires an extremely small time-step size to accurately capture the bubble growth rate as well as the temperature at the liquid-vapor interface, thus hindering the simulation of large-scale flash boiling spray cases due to the high computational cost associated with it. \par
Realizing the need for developing a simple correlation for bubble growth, a scaling law was first explored in this study to assess its ability to characterize bubble growth for superheated microdroplet cases under subatmospheric operating conditions. It was observed that the scaling law does not result in a perfect collapse of the scaled radius-time curves over the whole lifetime of the vapor bubble.\par
To derive a simplified solution of the modified RPE, a dimensional analysis was performed for different orders of magnitudes of Reynolds numbers. It was revealed that the nondimensional acceleration term possesses much lower significance in determining the bubble growth characteristics as the order of magnitude of the Re is decreased from $\mathcal{O}(10^{1})$ to $\mathcal{O}(10^{0})$. Thus, neglecting this term, a simple semi-analytical solution was derived for the bubble growth rate considering the bubble-bubble interactions and validated for different operating conditions against its numerical solution. The derived semi-analytical solution for bubble growth was found to provide a good approximation of the bubble radius as a function of nondimensional time. \par
A posteriori computational cost analysis for the single droplet cases also revealed that the proposed semi-analytical solver is significantly faster compared to the numerical solver; thus, it may be used in many practical applications including but not limited to flash-boiling spray simulations with millions of single droplets in automotive or cryogenic engines, where tracking of bubble growth phases in an individual droplet by solving the full RPE simultaneously with the energy balance at the liquid-vapor interface may be computationally very expensive.
\section*{Acknowledgements}
This work was performed as part of the Cluster of Excellence “The Fuel Science Center”, which is funded by the Deutsche Forschungsgemeinschaft (DFG, German Research Foundation) under Germany’s Excellence Strategy – Exzellenzcluster 2186
“The Fuel Science Center” ID: 390919832.
\appendix
\section{Previous scaling relationships and nondimensional analytical solutions}\label{appA}
\subsection{Scaling laws}
The first scaling law proposed by \cite{Prosperetti1978} based on the theory provided by \cite{Plesset1954} is given by \begin{equation}\label{Pros}
\tilde{\Rb}=\frac{C^{2} \Rb}{\Rc}, \quad \tilde{t}=D C^{2} t,
\end{equation}
and
\begin{equation} \label{Cconst}
C=\left(\frac{2 \sigma\left(\Tl\right) \alphal}{9\pi}\right)^{\frac{1}{2}}  \frac{\rhov \Lv}{\lambdal}\frac{\left\{\rhol\left[\Pv-\Pl\right]\right\}^{-\frac{1}{2}}}{\left(\Tl-\Tsat\left(\Pl\right)\right)}, \\  
\end{equation}
\begin{equation} \label{Dconst}
D=\frac{\left[\Pv-\Pl\right]^{\frac{3}{2}}}{2 \sigma\left(\Tl\right) \rhol^{\frac{1}{2}}},
\end{equation}
where `$\,\tilde{}\,$' implies nondimensional quantities. The physical parameters of \refe{Cconst} and \refe{Dconst} are evaluated at $\Tsat\left(\Pl\right)$, except $\sigma$. \par
The second scaling law proposed by \cite{Prosperetti1978} is given by 
\begin{equation}\label{Pros2}
\hat{\Rb}=\frac{\Rb}{\Rc}, \quad \hat{t}=Dt,
\end{equation}
where `$\,\hat{}\,$' denotes nondimensional quantities.
\subsection{Analytical solutions}
The nondimensional analytical solutions derived by \cite{Saha2021} are expressed as
\begin{equation}
    \frac{\text d\Rb^+}{\text dt^+}=\sqrt{\frac{2}{3}\left(\Pvplus-\Pgplus\right)\left\{1-\left(\frac{1}{\Rb^{+}}\right)^{3}\right\}-\frac{2}{\Rb^{+} \text{We}}\left\{1-\left(\frac{1}{\Rb^{+}}\right)^{2}\right\}} \quad\hspace{0.1cm}\text{for high Re},
\end{equation}
and
\begin{equation}
    \frac{\text d\Rb^+}{\text dt^+}=-\frac{4}{3\Rb^{+} \text{Re}} + \sqrt{\left(\frac{4}{3\Rb^{+} \text{Re}}\right)^{2}-\left\{\frac{4}{3\Rb^{+} \text{We}}-\frac{2}{3}\left(\Pvplus-\Pgplus\right)\right\}} \quad\hspace{0.1cm}\text{for low Re},
\end{equation}
\section{Derivation of the pressure force due to bubble-bubble interactions}\label{appB}
Let us consider a spherically symmetric superheated microdroplet consisting with multiple bubbles. The pressure force, $\Pinertia$, induced by the $i$-th bubble on the target bubble, located at the center of the droplet can be obtained by solving the continuity and momentum equations for an inviscid incompressible liquid flow~\citep{Mettin1997}. The one-dimensional momentum and continuity equations in spherical coordinates are given by:
\begin{equation}\label{Mom}
\frac{\partial\ul}{\partial t}+\ul \frac{\partial\ul}{\partial r}=-\frac{1}{\rhol} \frac{\partial \Pinertia}{\partial r},
\end{equation}
\begin{equation}\label{Cont1}
\frac{\partial}{\partial r}\left(\ul r^{2}\right)=0,
\end{equation}
where $\ul(r,t)$ describes the velocity in the liquid phase relative to the interface, $r$ denotes the radial coordinate with the origin at the droplet center, and $t$ is the time. Integrating the continuity equation (\refe{Cont1}) in radial direction yields
\begin{eqnarray}\label{Cont2}
\ul(r, t)=\frac{F(t)}{r^{2}}.
\end{eqnarray}
$F(t)$ can be obtained by prescribing the kinematic boundary condition at the bubble surface.\par
The vapor mass flow rate for a single bubble can be computed as 
\begin{equation}\label{mdotb1}
\dot m_\text{b}=4\pi \Rb^2\rho_{v}\frac{\mathrm d \Rb}{\mathrm d t}+\frac{4}{3}\pi\Rb^3\frac{\mathrm d\rho_v}{\mathrm dt},
\end{equation}
where $\rhov$ is the saturated vapor density at $\Tv$. Neglecting the second term on the right-hand side of \refe{mdotb1}, as it can be shown to be very small compared to the first term, $\dot m_b$ can be expressed as~\citep{Brennen2013}
\begin{equation}\label{mdotb}
\dot m_\text{b}=4\pi \Rb^2\rho_{v}\frac{\mathrm d \Rb}{\mathrm d t}.
\end{equation}
Equating \refe{mdotb} with the inward liquid mass flux yields
\begin{eqnarray}
\rhol u_\text{l,a}=\rhov\frac{\mathrm d \Rb}{\mathrm d t},\\
u_\text{l,a}=\frac{\rhov}{\rhol}\frac{\mathrm d \Rb}{\mathrm d t},
\end{eqnarray}
where subscript `a' represents absolute quantity and $\rhol$ the saturated liquid density at $\Tv$. Thus, the relative velocity of the liquid across the interface ($r=R_\text{b}$),
\begin{eqnarray}\label{uR1}
\ul(\Rb, t)=\frac{\mathrm d \Rb}{\mathrm d t}-\frac{\rhov\left(\Tv\right)}{\rhol} \frac{\mathrm d \Rb}{\mathrm d t}=\left[1-\frac{\rhov\left(\Tv\right)}{\rhol}\right] \frac{\mathrm d \Rb}{\mathrm d t}.
\end{eqnarray}
Substituting \refe{uR1} into \refe{Cont2}, $F(t)$ at the liquid/vapor interface can be obtained as
\begin{equation}\label{Ft}
F(t)=\left[1-\frac{\rho_{v}\left(\Tv\right)}{\rhol}\right] \Rb^{2} \frac{\mathrm d \Rb}{\mathrm d t}.
\end{equation}
Assuming $\rhov\ll\rhol$, \refe{Ft} becomes
\begin{equation}
F(t)=\Rb^{2} \frac{\mathrm d \Rb}{\mathrm d t}.
\end{equation}
Substituting $F(t)$ into \refe{Cont2} yields,
\begin{eqnarray}\label{uR2}
u(r, t)=\frac{\Rb^2}{r^2} \frac{\mathrm d \Rb}{\mathrm d t}.
\end{eqnarray}
Replacing $u(r,t)$ in \refe{Cont2} with \refe{uR2} yields
\begin{equation}\label{Mom2}
    \frac{\partial}{\partial t}\left(\frac{\Rb^2}{r^2} \frac{\mathrm d \Rb}{\mathrm d t}\right)+\left(\frac{\Rb^2}{r^2} \frac{\mathrm d \Rb}{\mathrm d t}\right) \frac{\partial}{\partial r}\left(\frac{\Rb^2}{r^2} \frac{\mathrm d \Rb}{\mathrm d t}\right)=-\frac{1}{\rhol} \frac{\partial\Pinertia}{\partial r}.
\end{equation}
Assuming $\Pinertia(r\to\infty,t)=0$ and integrating \refe{Mom2} along $r$, $\Pinertia$ can be expressed as
\begin{equation}\label{Pint}
    \Pinertia=\frac{\rhol}{r} \frac{\mathrm{d}}{\mathrm{d} t}\left(\Rb^2 \frac{\mathrm d \Rb}{\mathrm d t}\right)+O\left(\frac{1}{r^{4}}\right).
\end{equation}
Neglecting the higher order term in \refe{Pint} and summing over multiple bubbles, $\Pinertia$ finally becomes
\begin{equation}
    \Pinertia=\rhol\frac{\mathrm{d}}{\mathrm{d}t}\left(\sum\frac{\Rb^{2}}{\Ri} \frac{\mathrm{d} \Rb}{\mathrm{~d} t} \right).
\end{equation}
\section{Derivation of the nondimensional RPE}\label{appC}
The RPE for an isolated spherically symmetric vapor bubble in a homogeneous infinite liquid medium is given by
\begin{equation}\label{RPE11}
\begin{split}
\Pv-\Pg & = \rhol \left[\Rb \frac{\mathrm{d}^{2} \Rb}{\mathrm{~d} t^{2}}+\frac{3}{2}\left(\frac{\mathrm{d} \Rb}{\mathrm{~d} t}\right)^{2}\right]+\frac{4{\mul}}{\Rb} \frac{\mathrm{d} \Rb}{\mathrm{d} t}+\frac{2 \sigma}{\Rb}.
\end{split}
\end{equation}
The modified RPE for spherically symmetric vapor bubble in a single droplet of radius $\Rd$ considering bubble-bubble interactions is given by 
\begin{equation}\label{RPEMod}
\begin{split}
    \Pv-\Pg=\rhol\left[\Rb\left(1+2 \pi \Rd^{2} n \Rb\right) \frac{\mathrm{d}^{2} \Rb}{\mathrm{~d} t^{2}}+\left(\frac{3}{2}+4 \pi \Rd^{2} n \Rb\right)\left(\frac{\mathrm{d} \Rb}{\mathrm{~d} t}\right)^{2}\right]+\frac{4 \mul}{\Rb}\frac{\mathrm{d} \Rb}{\mathrm{~d} t}\\+4 \pi n \rhol \Rd \Rb^{2} \left(\frac{\mathrm{d} \Rd}{\mathrm{~d} t}\right) \left(\frac{\mathrm{d} \Rb}{\mathrm{~d} t}\right)+\frac{2\sigma}{\Rb}.
\end{split}
\end{equation}
Substituting the following relationships
\begin{equation}\label{eqn:scl_qnt}
\begin{split}
\Rbplus &=\frac{\Rb}{\Rc}, \quad \dot\Rbplus=\frac{\dot\Rb}{A}, \quad \tplus=\frac{t}{\tau}, \quad \Rdplus=\frac{\Rd}{\Rc}, \quad \Ndenplus=n\Rc^3;\\
\rholplus &=\frac{\rhol}{\rholzero}, \quad \Pplus =\frac{P}{\rholzero A^{2}}, \quad \sigmaplus =\frac{\sigma}{\sigmazero}, \quad \mulplus=\frac{\mul}{\mulzero},
\end{split}
\end{equation}
into \refe{RPE11} yields,
\begin{equation}\label{RPEnd1}
\begin{split}
\left(\Pvplus-\Pgplus\right)\rholzero A^2=\rholplus\Rbplus\rholzero A^2\frac{\mathrm{d}^{2}\Rbplus}{\mathrm{d}\tplustwo}+\rholplus\rholzero A^2\frac{3}{2}\left(\frac{\mathrm{d} \Rbplus}{\mathrm{~d}\tplus}\right)^{2}+\frac{4\mulplus\mulzero A}{\Rbplus\Rc}\frac{\mathrm{d}\Rbplus}{\mathrm{d}\tplus}+\frac{2 \sigma^+\sigma_0}{\Rbplus\Rc}.
\end{split}
\end{equation}
Dividing both sides of \refe{RPEnd1} by $\rholzero A^2$ one obtains
\begin{equation}\label{RPEnd2}
\begin{split}
\Pvplus-\Pgplus=\rholplus\Rbplus\frac{\mathrm{d}^{2}\Rbplus}{\mathrm{d}\tplustwo}+\rholplus\frac{3}{2}\left(\frac{\mathrm{d} \Rbplus}{\mathrm{~d}\tplus}\right)^{2}+\frac{4\mulplus\mulzero }{\Rbplus\rholzero A\Rc}\frac{\mathrm{d}\Rbplus}{\mathrm{d}\tplus}+\frac{2 \sigma^+\sigma_0}{\Rbplus\rholzero A^2\Rc}.
\end{split}
\end{equation}
Defining $\text{Re}=\frac{\rholzero A\Rc}{\mulzero}$ and $\text{We}=\frac{\rholzero A^2\Rc}{\sigmazero}$, the nondimensional RPE for an isolated vapor bubble in an infinite liquid medium without bubble-bubble interactions is expressed as 
\begin{equation}\label{RPEnd3}
\begin{split}
\Pvplus-\Pgplus=\rholplus\Rbplus\frac{\mathrm{d}^{2}\Rbplus}{\mathrm{d}\tplustwo}+\rholplus\frac{3}{2}\left(\frac{\mathrm{d} \Rbplus}{\mathrm{~d}\tplus}\right)^{2}+\frac{4\mulplus\mulzero}{\Rbplus \text{Re}}\frac{\mathrm{d}\Rbplus}{\mathrm{d}\tplus}+\frac{2 \sigma^+}{\Rbplus \text{We}}.
\end{split}
\end{equation}
Substituting the scaling relations shown in \refe{eqn:scl_qnt} into \refe{RPEMod} yields
\begin{equation}\label{RPEnd4}
\begin{split}
\left(\Pvplus-\Pgplus\right)\rholzero A^2=\rholplus\Rbplus\rholzero A^2\left(1+2 \pi\Rdplustwo\Ndenplus\Rbplus\right)\frac{\mathrm{d}^{2}\Rbplus}{\mathrm{d}\tplustwo}+\rholplus\rholzero A^2\left(\frac{3}{2}+4 \pi\Rdplustwo\Ndenplus\Rbplus\right)\left(\frac{\mathrm{d} \Rbplus}{\mathrm{~d}\tplus}\right)^{2}+\frac{4\mulplus\mulzero A}{\Rbplus\Rc}\frac{\mathrm{d}\Rbplus}{\mathrm{d}\tplus}\\+4\pi \Ndenplus\rholplus\Rdplus\Rbplustwo\rholzero A^2\left(\frac{\mathrm{d}\Rdplus}{\mathrm{d}\tplus}\right)\left(\frac{\mathrm{d} \Rbplus}{\mathrm{d} \tplus}\right)+\frac{2 \sigma^+\sigma_0}{\Rbplus\Rc}.
\end{split}
\end{equation}
Dividing both sides by $\rholzero A^2$, \refe{RPEnd4} becomes
\begin{equation}\label{RPEnd2}
    \begin{split}
      \Pvplus-\Pgplus=\rholplus\Rbplus\left(1+2 \pi\Rdplustwo\Ndenplus\Rbplus\right)\frac{\mathrm{d}^{2}\Rbplus}{\mathrm{d}\tplustwo}+\rholplus\left(\frac{3}{2}+4 \pi\Rdplustwo\Ndenplus\Rbplus\right)\left(\frac{\mathrm{d} \Rbplus}{\mathrm{~d}\tplus}\right)^{2}+\frac{4\mulplus\mulzero }{\Rbplus\rholzero A\Rc}\frac{\mathrm{d}\Rbplus}{\mathrm{d}\tplus}\\+4\pi \Ndenplus\rholplus\Rdplus\Rbplustwo\left(\frac{\mathrm{d}\Rdplus}{\mathrm{d}\tplus}\right)\left(\frac{\mathrm{d} \Rbplus}{\mathrm{d} \tplus}\right)+\frac{2 \sigma^+\sigma_0}{\Rbplus\rholzero A^2\Rc}.
   \end{split}
\end{equation}
Substituting $\text{Re}=\frac{\rholzero A\Rc}{\mulzero}$ and $\text{We}=\frac{\rholzero A^2\Rc}{\sigmazero}$, the nondimensional RPE of a spherically symmetric vapor bubble in a single droplet configuration considering bubble-bubble interactions is given by
\begin{equation}\label{RPE_ND}
    \begin{split}
      \Pvplus-\Pgplus=\rholplus\Rbplus\left(1+2 \pi\Rdplustwo\Ndenplus\Rbplus\right)\frac{\mathrm{d}^{2}\Rbplus}{\mathrm{d}\tplustwo}+\rholplus\left(\frac{3}{2}+4 \pi\Rdplustwo\Ndenplus\Rbplus\right)\left(\frac{\mathrm{d} \Rbplus}{\mathrm{~d}\tplus}\right)^{2}+\frac{4\mulplus}{\Rbplus \text{Re}}\frac{\mathrm{d}\Rbplus}{\mathrm{d}\tplus}\\+4\pi \Ndenplus\rholplus\Rdplus\Rbplustwo\left(\frac{\mathrm{d}\Rdplus}{\mathrm{d}\tplus}\right)\left(\frac{\mathrm{d} \Rbplus}{\mathrm{d} \tplus}\right)+\frac{2 \sigma^+}{\Rbplus \text{We}}.
   \end{split}
\end{equation}
\section{Model validation results at single isolated bubble level}\label{appD}
\begin{figure}[!h]
\centering
\includegraphics[width=420pt]{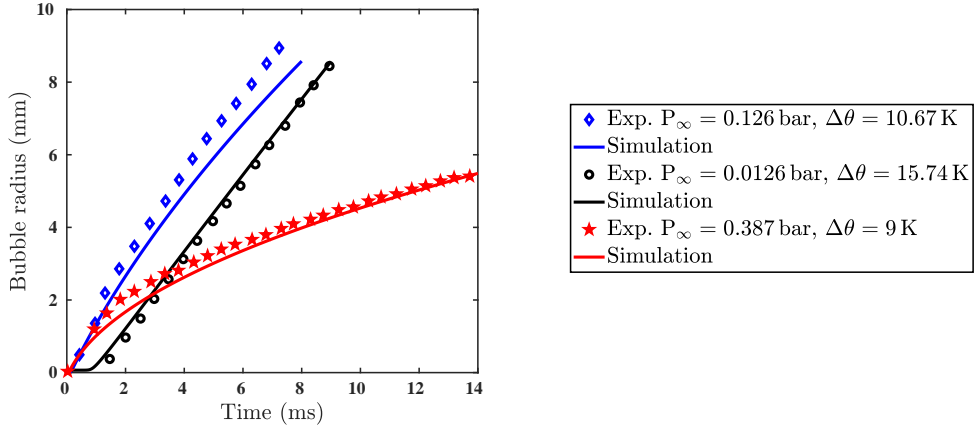}  
\caption{Single isolated vapor bubble growth behavior in superheated liquid water for various pressures and temperatures.}
\label{WaterVald}
\end{figure}

\reff{WaterVald} shows the validation of the bubble growth model for a single isolated vapor bubble growth in a homogeneous infinite superheated liquid water medium. The reader is referred to \cite{Saha2022} for more details about the simulation setup considered for this validation. 
\section{Limitations of the numerical solver for DME}\label{appE}
\begin{figure}
\centering
\includegraphics[width=420pt]{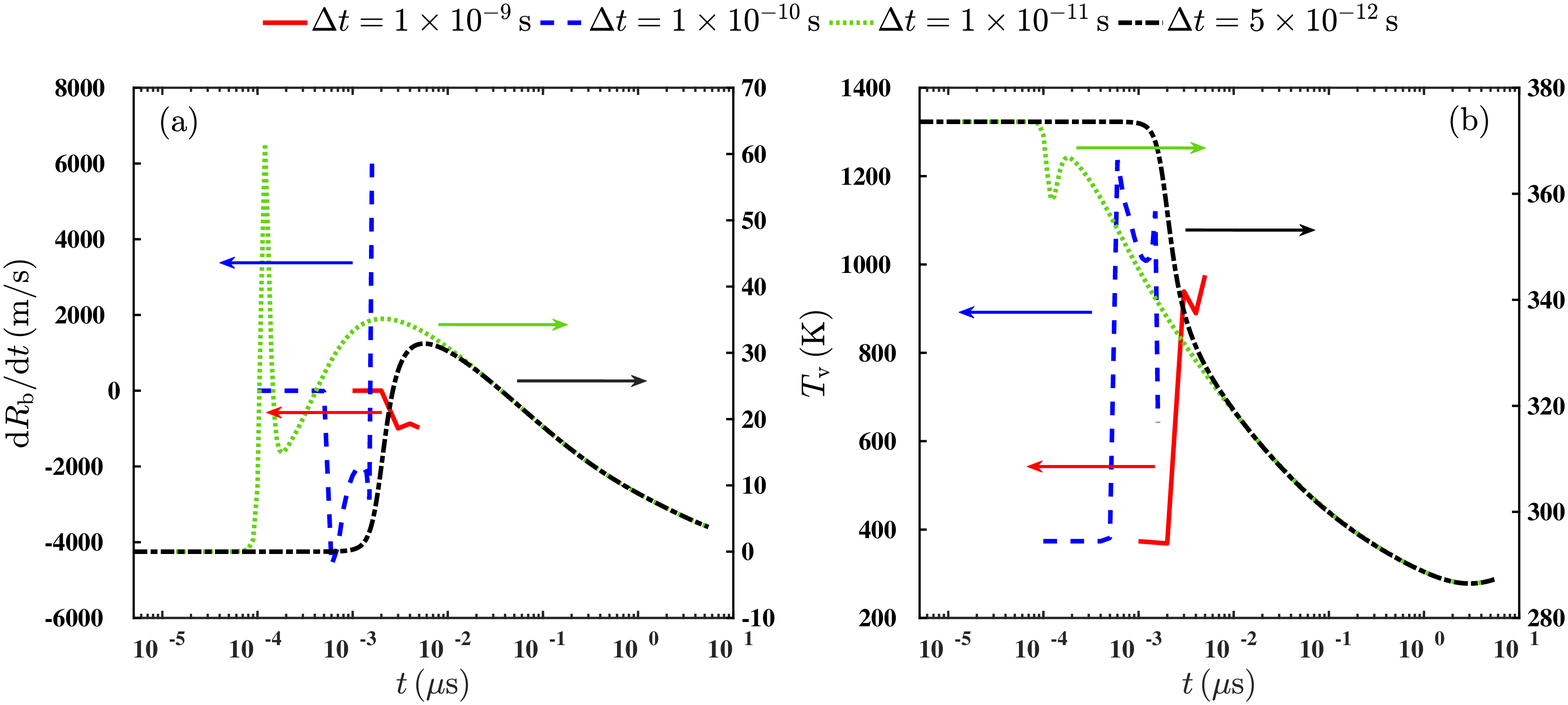}  
\vspace{-0.4 in}
\includegraphics[width=420pt]{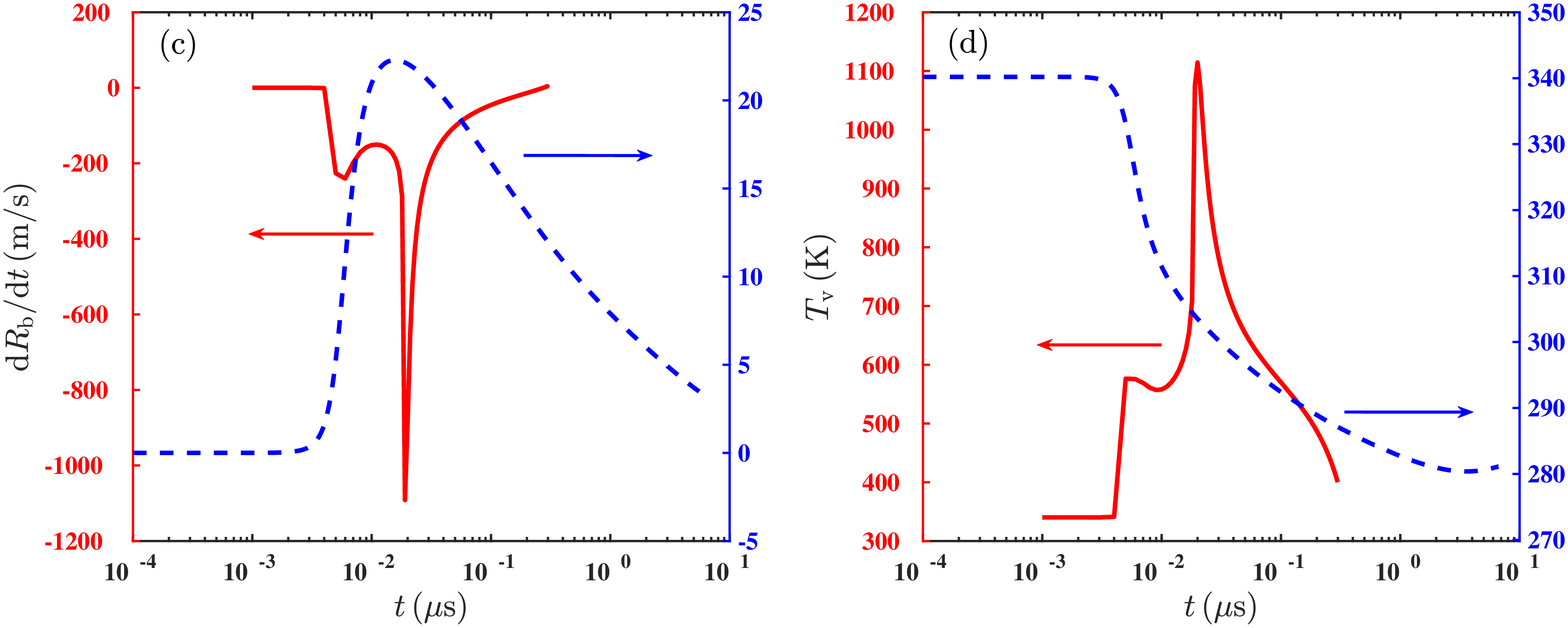}  
\includegraphics[width=420pt]{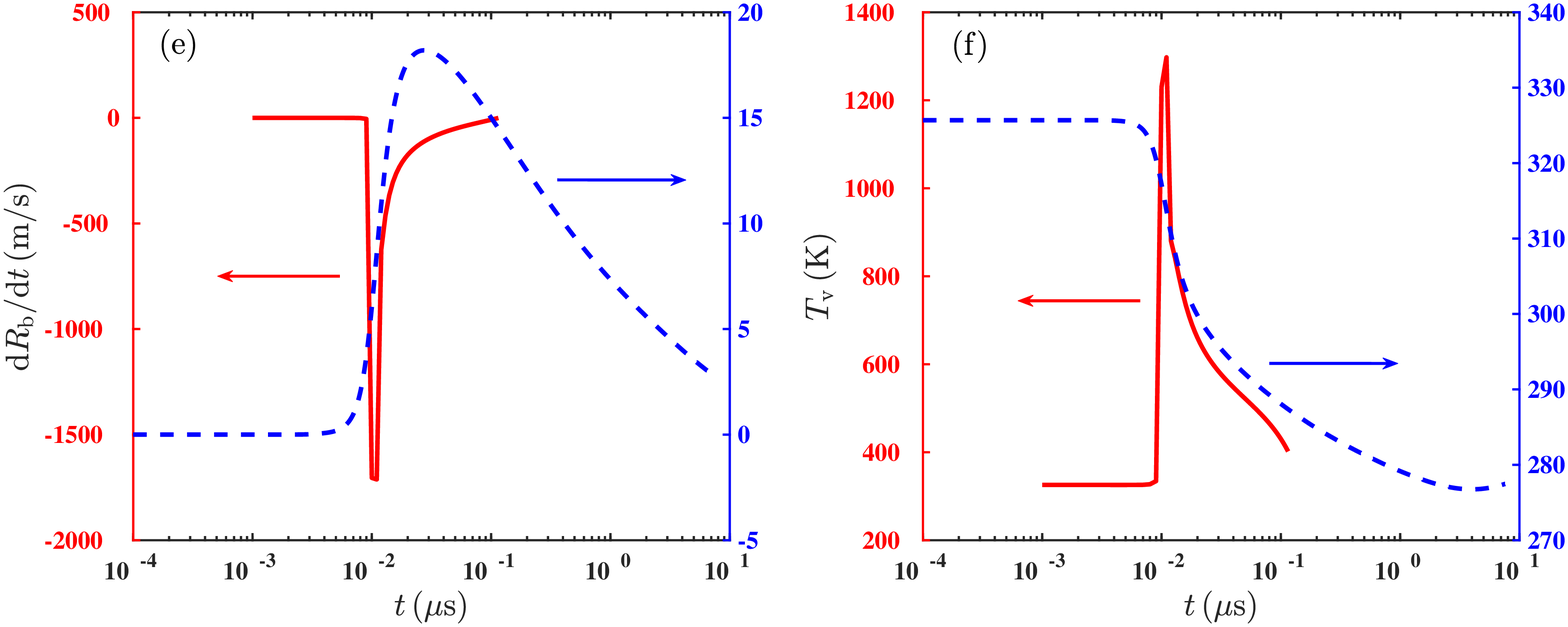}  
\vspace{-0.5 in}
\caption{Bubble growth rate (left column) and vapor temperature (right column) variation for DME microdroplets under various operating conditions and time-step sizes considering the bubble-bubble interactions. Subfigures (a) $\&$ (b), (c) $\&$ (d), and (e) $\&$ (f) depict the cases `DM-150', `DM-100', and `DM-80', respectively.}
\label{Saha8910wB}
\end{figure}

The numerical issue with the time-step size is also demonstrated using another highly volatile cryogenic e-fuel microdroplets of DME in \reff{Saha8910wB} for $\Delta\theta=150$ K (Case `DM-150'), $\Delta\theta=100$ K (Case `DM-100'), and $\Delta\theta=80$ K (Case `DM-80'), with bubble-bubble interactions. The simulation parameters are listed in \reft{LMCases2}. As described for $\OMEa$ microdroplets, the smaller time-step size is also required for the accurate prediction of vapor bubble growth in DME microdroplets. For a very high superheating degree of $\Delta\theta=150$ K, the time-step size needs to be further reduced to the order of picoseconds $\left(\mathcal{O}\left(10^{-12}\,\text{s}\right)\right)$ to avoid unphysical prediction of bubble growth and vapor temperature.
\begin{table}[h]
  \begin{center}
\def~{\hphantom{0}}
  \begin{tabular}{l@{\quad}l@{\quad}c@{\quad}l@{\quad}r@{\quad}r}
  \toprule
      \textbf{Case} & \hskip0.5cm\textbf{$\Pl \left(\mathrm{bar}\right)$} & \hskip0.5cm\textbf{$\Td \left(\mathrm K\right)$}          & \hskip0.5cm\textbf{$\Delta\theta \left(\mathrm K\right)$}   \\ [3pt]
  \midrule
DM-150           & \hskip0.5cm 0.3                                                       & \hskip0.5cm 373.58       & \hskip0.5cm 150               \\ [3pt]
DM-100           & \hskip0.5cm 0.7                                                       & \hskip0.5cm 340.2       & \hskip0.5cm 100               \\ [3pt]
DM-80           & \hskip0.5cm 0.9                                                       & \hskip0.5cm 325.68      & \hskip0.5cm 80                                                                           \\
  \bottomrule
  \end{tabular}
  \caption{Simulation test cases of superheated DME microdroplets for illustrating the limitations of the numerical solver.}
  \label{LMCases2}
  \end{center}
\end{table}

\appendix

\bibliography{references}






\end{document}